\documentclass[english,aps,prb,showpacs,superscriptaddress,floats,amsmath,amssymb,floatfix,nobalancelastpage,twocolumn]{revtex4-1}
\usepackage[bookmarks=true,colorlinks,linkcolor=blue,urlcolor=blue,citecolor=blue]{hyperref}
\usepackage[T1]{fontenc}
\usepackage[latin9]{inputenc}
\usepackage{color}
\usepackage{amsmath}
\usepackage{amssymb}
\usepackage{graphicx}

\makeatletter
\usepackage{accents}

\makeatother

\usepackage{babel}
\begin{document}
\global\long\def\ptilde#1{\overset{{\scriptscriptstyle {(\sim)}}}{#1}}%

\title{Spinless fermions in a $\mathbb{Z}_{2}$ gauge theory on the triangular
ladder}
\author{Wolfram Brenig}
\affiliation{Institute for Theoretical Physics, Technical University Braunschweig,
D-38106 Braunschweig, Germany}
\begin{abstract}
A study of spinless matter fermions coupled to a constrained $\mathbb{Z}_{2}$
lattice gauge theory on a triangular ladder is presented. The triangular
unit cell and the ladder geometry strongly modify the physics, as
compared to previous analysis on the square lattice \citep{Borla2021}.
In the static case, the even and odd gauge theories for the empty
and filled ladder are identical. The gauge field dynamics due to the
electric coupling is drastically influenced by the absence of periodic
boundary conditions, rendering the deconfinement-confinement process
a crossover in general and a quantum phase transition (QPT) only for
decorated couplings. At finite doping and in the static case, a staggered
flux insulator at half filling and vanishing magnetic energy competes
with a uniform flux metal at elevated magnetic energy. As for the
square lattice, a single QPT into a confined fermionic dimer gas is
found versus electric coupling. Dimer resonances in the confined phase
are however a second order process only, likely reducing the tendency
to phase separate for large magnetic energy. The results obtained
employ a mapping to a pure spin model of $\mathbb{Z}_{2}$ gauge-invariant
moments, adapted from the square lattice, and density matrix renormalization
group calculations thereof for numerical purposes. Global scans of
the quantum phases in the intermediate coupling regime are provided.
\end{abstract}
\maketitle

\section{Introduction}

Paradigmatic models of frustrated quantum magnetism can be viewed
as gauge field theories, featuring topological phases with emergent
non-local excitations of anyonic statistics \citep{Henley2010,Castelnovo2012,Savary2016,Sachdev2018}.
A celebrated example is Kitaev's toric code \citep{Kitaev2003}. If
coupled to a magnetic field and without gauge charges, it relates
to Wegner's $\mathbb{Z}_{2}$ gauge theory \citep{Wegner1971}, which
is dual \citep{Kramers1941,Kogut1979} to the transverse field Ising
model. This so called Ising gauge theory (IGT) is well known to exhibit
a deconfinement-confinement transition in terms of Wegner-Wilson loops
\citep{Wegner1971}. It is exactly this transition which is not characterized
by a local Ginzburg-Landau order parameter, but rather it separates
a topologically ordered \citep{Kitaev2003,Wen1991}, i.e., deconfined,
from a trivial, i.e., confined phase. Additional examples of current
interest involve, e.g., the $U(1)$ gauge theories of hard core dimers
in three dimensions, or spin ice in easy axis pyrochlore magnets and
their Coulomb phase \citep{Huse2003,Hermele2004,Savary2012,Henley2010}.

Coupling of gauge fields to matter arises naturally in most slave-particle,
or parton descriptions of quantum magnets, where the original spin
degrees of freedom are fractionalized in terms of Dirac fermions \citep{Abrikosov1965,Baskaran1987,Affleck1988a},
Majorana fermions \citep{Kitaev2006}, or bosons \citep{Schwinger1965,Arovas1988,Read1991}.
Depending on extensively classified sets of mean-field starting points
\citep{Wen2002,Wang2006,Messio2013}, restoring the original from
the enlarged, fractionalized Hilbert spaces, induces a coupling of
the parton matter with lattice gauge fields, leading to theories of
$SU(2)$, $U(1)$, $\mathbb{Z}_{2}$, and more exotic symmetries.\textcolor{blue}{{}
}This concept has been of interest early on, for local moment Anderson
impurities and lattices \citep{Read1983}, Heisenberg antiferromagnets
\citep{Affleck1988,Elbio1988}, and Hubbard models \citep{Kim2006,Hermele2007,Sachdev2009},
comprising primarily $U(1)$ and $SU(2)$ gauge theories.

For $\mathbb{Z}_{2}$ gauge theories, undoubtedly, Kitaev's anisotropic
Ising-exchange Hamiltonian on the honeycomb lattice is of great current interest
\citep{Kitaev2006}. It is one of the few models, in which a $\mathbb{Z}_{2}$
quantum spin liquid (QSL) can exactly be shown to exist, following the route of
fractionalizing spin degrees of freedom, namely in terms of mobile Majorana
fermions coupled to a static $\mathbb{Z}_{2}$ gauge field
\citep{Kitaev2006,Feng2007,Chen2008,Nussinov2009,Mandal2012}.  Here, gauge flux
dynamics can be induced by external magnetic fields \citep{Kitaev2006} and
non-Kitaev exchange \citep{Zhang2021,Joy2021}. Extensions including orbital degrees
of freedom have been considered \citep{Yao2011,Seifert2020}.  The high-energy
properties of $\alpha$-RuCl$_{3}$ \citep{Plumb2014} may be a territory to look for
this physics, even though the low-energy behavior is dominated by magnetic order
\citep{Trebst2017,Winter2017,Janssen2019}.

Early on, the coupling of $\mathbb{Z}_{2}$ gauge fields to matter
was also considered in a broader context, using Ising-like scalar
Higgs matter-fields \citep{Fradkin1979}. In that setting, the phases
of Wegner's $\mathbb{Z}_{2}$ gauge field theory were shown to persist,
and an additional Higgs regime was found to belong to the confined
phase. This is consistent with quantum Monte Carlo analysis \citep{Trebst2007,Tupitsyn2010}.
Following the discovery of the cuprate superconductors, $\mathbb{Z}_{2}$
gauge fields coupled to spin-charge separated matter have also been
invoked to analyze strongly correlated electron systems, e.g. \citep{Senthil2000}.
Lately, non-Fermi liquid behavior has been proposed for so called
orthogonal metals \citep{Nandkishore2012}, comprising an IGT for
a slave-spin representation of fermions. Finally, ultracold atomic
gas setups have realized unit cells of the toric code very recently
\citep{Homeier2021}.

In line with these general developments, lattice IGTs, constrained
or unconstrained, and minimally coupled to either free fermions \citep{Zhong2013,Assaad2016,Gazit2017,Prosko2017,Konig2020,Cuadra2020,Borla2020,Borla2021a,Borla2021},
the Hubbard model \citep{Gazit2018,Gazit2020}, or composite fermions
\citep{Chen2020}, are currently experiencing an upsurge of attention.
The phases of these models are very diverse. They can host non-Fermi-liquids
of the orthogonal metal (OM) and semimetal (OSM) type, and may allow
for Fermi-surface reconstruction without symmetry breaking - all of
which arises from the dressing of the fermions by the $\mathbb{Z}_{2}$
gauge field. They incorporate attractive interactions between the
fermions from the $\mathbb{Z}_{2}$ gauge field, which, depending
on the strength of the confinement, i.e., the string tension, can
lead to BCS superconductors or BEC superfluids, and corresponding
QPTs between them. In the presence of finite Hubbard repulsion, QPTs
from OSMs into antiferromagnetism (AFM) can occur versus increasing
Hubbard repulsion but also versus string tension. The latter case
is under intense debate as to whether the gapping of the fermionic
spectrum and the confinement are a two stage, or single transition.
Recently, this may have been settled in favor of a single transition
involving $SO(5)$ symmetry \citep{Gazit2018}.

While spinful fermions allow for magnetic order, spinless fermions
or Majorana combinations thereof, are also among the parton matter
which has been coupled to lattice IGTs in one \citep{Borla2020,Borla2021a},
and two \citep{Borla2021} dimensions (1D,2D). In 2D, many similarities
arise with theories comprising spinful fermions. In particular, Fermi-surface
reconstruction in combination with a topological transition between
differing flux-phases is found in the deconfined region. Additionally,
a QPT into a confined phase of a dimer Mott-state is observed, which
phase separates for sufficiently large flux energies.

Naturally, lattice IGTs are not only sensitive to the dimension, but
also to the underlying lattice structure, where hypercubic geometries
are the conventional playground. In the present work, a step is taken
away from that, by considering spinless fermions coupled to a lattice
IGT on a \emph{triangular ladder}. Various aspects of its quantum
phases are studied versus the electric and magnetic energies, as well
as the fermionic filling and compared to findings on the 2D square
lattice \citep{Borla2021}. It is shown that the ladder generates
a significantly modified picture.

The structure of the paper is as follows: In Sec. \ref{sec:model}
the model is described, and in Sec. \ref{sec:SpinReps} it is reformulated
in terms of a spin-only Hamiltonian. Sec. \ref{sec:Results} presents
the results in various limiting cases, comprising the pure gauge theories
in Subsec. \ref{subsec::Pure}, the static case in Subsec. \ref{subsec:stagphi},
the strongly confined limit in Subsec. \ref{subsec:largeh}, the transition
into confinement at half filling in Subsec. \ref{subsec:stagddwtrans},
and finally a scan of quantum phases over a range of all-intermediate
parameters in Subsec. \ref{subsec:scan}. In Sec. \ref{sec:Conclusion}
conclusions are given. Appendix \ref{app:A} contains technical details
of a mapping to the spin-only Hamiltonian.

\begin{figure}[tb]
\begin{centering}
\includegraphics[width=0.6\columnwidth]{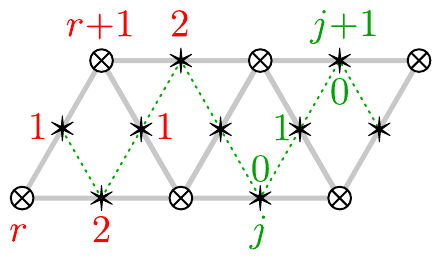}
\par\end{centering}
\caption{\label{fig:lat}(Color online) Original lattice (open crossed circle
on gray solid) with sites $r$. Dual lattice (filled stars, dashed
green, green labels) with sites $b=j,i$, and $i=0,1$. Dual lattice
sites are also labeled by $b=r,i$ (red labels), with $i=1(2)$, for
rungs (legs).}
\end{figure}

\section{The Model \label{sec:model}}

This work deals with spinless fermions coupled to a constrained $\mathbb{Z}_{2}$
Ising gauge theory (SFIGT) on the triangular ladder depicted in Fig.
\ref{fig:lat}. Prior to defining the model, nomenclature for the
lattice is introduced in this figure. It shows the original lattice
and its dual. Sites of the original lattice are labeled by $r$. In
principle, this should be expressed in terms of the triangular basis.
For simplicity however, and because of the quasi 1D geometry, $r$
is enumerated using $r\in\mathbb{Z}$. Sites on the dual lattice are
either labeled by tuples for the corresponding bonds $b=j,i$, using
$j\in\mathbb{Z}$ and $i=0,1$, or in terms of the original lattice
by the tuples $b=r,i$, with $i=1$(2) for rungs(legs). Finally, $r-i,i\equiv r,-i$
is used.

With the preceding, the gauge theory, coupled to the spinless fermionic
matter is
\begin{equation}
H=H_{c}+H_{g}\,.\label{eq:1}
\end{equation}
The matter is modeled by
\begin{equation}
H_{c}=-\sum_{r,i=1,2}(t_{i}c_{r+i}^{\dagger}\sigma_{r,i}^{z}c_{r}^{\phantom{\dagger}}+h.c.)-\mu\sum_{r}n_{r}\,,\label{eq:2}
\end{equation}
where $t_{i}$ are nearest(next-nearest) neighbor hopping matrix elements
for $i=1(2)$. The fermions are (created) destroyed by $c_{r}^{(\dagger)}$
on sites $r$. $\sigma_{r,i}^{\alpha}$, with $\alpha=x,y,z$ are
Pauli matrices which reside on the sites $r,i$ of the dual lattice,
and $\sigma_{r,i}^{z}$ is the equivalent of the Peierls factor for
the $\mathbb{Z}_{2}$ gauge theory. $\mu$ is the chemical potential
and $n_{r}=c_{r}^{\dagger}c_{r}^{\phantom{\dagger}}$ is the fermion
number on site $r$, i.e., the physical charge.

\begin{figure}[tb]
\begin{centering}
\includegraphics[width=0.8\columnwidth]{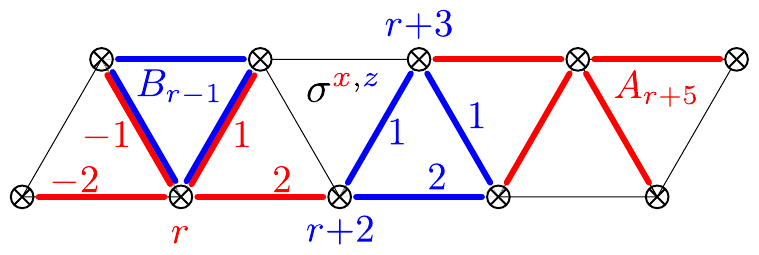}
\par\end{centering}
\caption{\label{fig:Z2ops}(Color online) Examples of star (red) and plaquette
(blue) operators, $A_{r}$ and $B_{r}$, from Eqs. (\ref{eq:4}) and
(\ref{eq:5}). Red (blue) bonds refer to $\sigma^{x}$($\sigma^{z}$)
operators on dual lattice sites, $r,i$ with $i=1,2$, of bond.}
\end{figure}

The constrained $\mathbb{Z}_{2}$ Ising gauge theory \citep{Wegner1971,Kogut1979}
on the triangular-ladder is given by
\begin{equation}
H_{g}=-J\sum_{r}\prod_{b\in\vartriangle_{r}(\triangledown_{r})}\sigma_{b}^{z}-\sum_{r,i=1,2}h_{r,i}\sigma_{r,i}^{x}\,;\hphantom{aa}G_{r}=1\label{eq:3}
\end{equation}
where the up(down)-triangularly shaped plaquettes $P_{r}=\vartriangle_{r}(\triangledown_{r})$
reside on the blue links, shown in Fig. \ref{fig:Z2ops}, and refer
to the sets of dual sites $b=\{(r,1),$ $(r,2),$ $(r+1,1)\}$ both,
for up- and down-plaquettes. The first term in Eq. (\ref{eq:3}) is
the magnetic field energy of the $\mathbb{Z}_{2}$ gauge theory, with
magnetic coupling constant $J$ and magnetic flux, or plaquette operator
\begin{equation}
B_{r}=\prod_{b\in P_{r}}\sigma_{b}^{z}.\label{eq:4}
\end{equation}
Since $B_{r}^{2}=1$, the flux has eigenvalues $\pm1$. The second
term is the electric field energy, where $h_{r,i}$ is the electric
coupling constant and $\sigma_{r,i}^{x}$ is the electric field operator\emph{
}\citep{Note1}. Since $(\sigma_{r,i}^{x})^{2}=1$, the electric field
has eigenvalues $\pm1$.

At this point, Eq. (\ref{eq:3}) allows for electric couplings, different
on the legs and rungs of the ladder. In the following \emph{ladder}-coupling
implies $h_{r,i}=h$, i.e., the electric field energy is identical
on the rungs and legs of the ladder, while \emph{chain}-coupling means
$h_{r,1}=h$ and $h_{r,2}=0$, i.e., the electric field energies exist
only on the chain formed by the rungs of the ladder. These different
couplings will play a role only in Subsec. \ref{subsec::Pure}. All
other results will be obtained using ladder-coupling.

The local $\mathbb{Z}_{2}$ gauge invariance of $H$ is encoded in
the corresponding generator
\begin{equation}
G_{r}=(-)^{n_{r}}\prod_{b\in S_{r}}\sigma_{b}^{x}\equiv(-)^{n_{r}}A_{r},\label{eq:5}
\end{equation}
where the squashed 'stars' $S_{r}$ refer to the set of dual sites
$b=\{(r,-2),(r,-1),(r,1),(r,2)\}$, both, for $r$ on the upper and
lower leg. These reside on the red links in Fig. \ref{fig:Z2ops}
and $A_{r}$ is the star operator. As for $\mathbb{Z}_{2}$ gauge
theories on square lattices, stars and plaquettes either share two,
or no dual lattice sites, i.e., star and plaquette operators commute
$[A_{r},B_{r'}]=0$, $\forall r,r'$. Therefore, $G_{r}$ is indeed
a symmetry $[H,G_{r}]=0$.

Conservation of $G_{r}$ is the $\mathbb{Z}_{2}$ version of Gau{\ss}'s
law. The eigenvalues of $G_{r}$ are the vacuum charges of the gauge
theory. Since $G_{r}^{+}G_{r}^{\phantom{+}}=G_{r}^{2}=1$, these can
be $\pm1$. As stated in Eq. (\ref{eq:3}), a homogeneous gauge vacuum
of $G_{r}=1$, $\forall r$ is used in the present work. This is the
so called even gauge theory, as compared to the odd one, for which
$G_{r}=-1$, $\forall r$. Fixing the vacuum charge per site defines
the notion of a constrained gauge theory -- as opposed to an unconstrained
one, where all values of gauge charges per site are allowed. For $G_{r}=1$,
valid configurations of the physical charge and electric field are
such that on each site, the total of the number of fermions and the
number of $\sigma_{b}^{x}=-1$ links on that site has to be $0\mod2$.
Such configurations are exemplified in Fig. \ref{fig:Z2ops}.

\begin{figure}[tb]
\begin{centering}
\includegraphics[width=0.8\columnwidth]{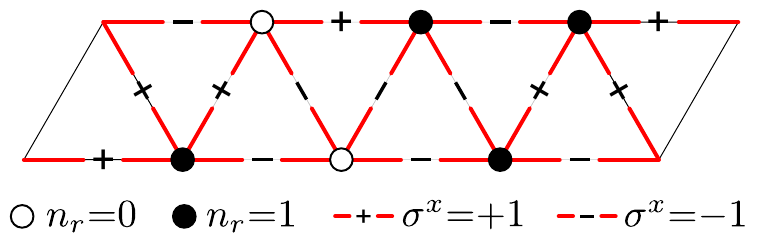}
\par\end{centering}
\caption{\label{fig:gauss}(Color online) Typical $\mathbb{Z}_{2}$, Gau{\ss} law
abiding configuration with $G_{r}=1$ of physical charges $n_{r}$
(solid and open black circles) and electric field $\sigma^{x}$ (red
$\pm$ bonds).}
\end{figure}

Bonds with $\sigma_{b}^{x}=-1$ are called electric strings. The constraint
and Gau{\ss}'s law force the number of fermions in any microcanonical
state to be even, since at any site $r^{\prime}$, at which a string
terminates which has been emitted by a fermion inserted at site $r$
previously, the fermion parity must change a second time. For $h>0$,
strings are energetically expensive with a potential increasing linearly
in the string length. In turn, pairs of fermions attract each other
in that case.

Next, several symmetries relevant for model (\ref{eq:1}) and its
operators are collected. All of them have been listed in the literature
\citep{Borla2021,Sachdev2018,Prosko2017,Gazit2018}. First, the action
of the $\mathbb{Z}_{2}$ generator on the fermions is $G_{r}c_{r}^{(\dagger)}G_{r}=-c_{r}^{(\dagger)}$,
i.e., the original fermions are not gauge-invariant. Similarly, $G_{r}\sigma_{b}^{y(z)}G_{r}=\eta\sigma_{b}^{y(z)}$,
where $\eta=-1$ if $b\in S_{r}$, otherwise $\eta=1$. Second, both,
the Hamiltonian and Gau{\ss} law are invariant under time inversion, which
is the identity for all spinless fermion creation (destruction) operators
and Pauli matrices, except for $\sigma_{b}^{y}$, which under complex
conjugation changes sign, i.e., $\sigma_{b}^{y}\rightarrow-\sigma_{b}^{y}$.
As compared to versions of the model (\ref{eq:1}) on bipartite lattices
\citep{Borla2021,Prosko2017,Gazit2018}, the fermionic matter of Eq.
(\ref{eq:2}) on the triangular ladder is not particle-hole symmetric,
at any $\mu$. Yet, the transformation $c_{r}^{\dagger}\rightarrow c_{r}$
maps $H_{c}(t_{1},t_{2},\mu)\rightarrow H_{c}(-t_{1},-t_{2},-\mu)$
and $G_{r}\rightarrow-G_{r}$. I.e., the complete model has even and
odd $\mathbb{Z}_{2}$ theories related by flipping the signs of all
parameters of the fermionic matter. The remainder of this work focuses
on $t_{1,2}$, $J$, and $h$ \emph{all} positive.

\section{Spin chain representation\label{sec:SpinReps}}

In App. \ref{app:A}, technical details of a mapping of the SFIGT
on the triangular ladder to a pure spin model with only two sets of
spin operators, $(X,Y,Z)_{r,i}$ per triangle are described. Variants
of this have already been used for 1D \citep{Borla2020,Borla2021a,Cobanera2013,Radicevic2018}
and 2D \citep{Borla2021} systems. The new spins are gauge-invariant,
and the pure spin model acts only in the sector of zero gauge charge
by construction. Since the new spins are labeled by the dual lattice
$(r,i=1,2)$ only, the transformed model can also be viewed as a 1D
spin chain with two sites per unit cell by using the notation from
Fig. \ref{fig:lat} with $(r,i=1,2)\rightarrow(j,i=0,1)$. In terms
of this chain notation, the transformed Hamiltonian in terms of $(X,Y,Z)_{j,i=0,1}$
reads
\begin{align}
H= & \sum_{j}H_{j}\,,\label{eq:6a}\\
H_{j}= & \big\{(-\frac{1}{2})\big[\,t_{1}\,(Z_{j,1}X_{j+1,0}\nonumber \\
 & \phantom{aaaaa}-Z_{j,1}X_{j-1,0}X_{j-1,1}X_{j,0}X_{j+1,1}X_{j+2,0})\nonumber \\
 & +\,t_{2}\,(Z_{j,0}X_{j,1}\nonumber \\
 & \phantom{aaaa}-Z_{j,0}X_{j-2,0}X_{j-2,1}X_{j-1,1}X_{j+1,1}X_{j+2,0})\big]\nonumber \\
 & -\frac{\mu}{2}(1-X_{j,0}X_{j,1}X_{j+1,0}X_{j+2,0})\nonumber \\
 & -JX_{j,0}Y_{j,1}Y_{j+1,0}Z_{j+1,1}\nonumber \\
 & -h_{j,0}X_{j,0}-h_{j,1}X_{j,1}\,,\label{eq:7a}
\end{align}
where $h_{r,1(2)}\equiv h_{j,1(0)}$. As compared to the original
model, comprising fermions, the reformulation Eq. (\ref{eq:6a}) has
the advantage that it allows for numerical calculations, using, e.g.,
DMRG, with a local Hilbert space reduced by a factor of $2$ and no
gauge constraint to be enforced aside. For more information on the
mapping, App. \ref{app:A} should be consulted.

\section{Results\label{sec:Results}}

In the following subsections, several limiting cases of the SFIGT
are considered. As has been shown in refs. \citep{Assaad2016,Gazit2017,Gazit2018},
in \citep{Prosko2017}, and in \citep{Borla2021}, this allows to
draw a qualitatively and quantitatively rather complete picture of
substantial regions of the quantum phase diagram. The order of the
discussion follows closely the one considered in ref. \citep{Borla2021}.
Despite this, the resulting behavior of the SFIGT in the present study
will deviate significantly from that reference.

\subsection{$\mu\rightarrow{-}({+})\infty\,$: Even(odd) pure $\mathbb{Z}_{2}$
gauge theory \label{subsec::Pure}}

For $\mu\rightarrow{-}({+})\infty$, the fermion sites are strictly
empty (occupied). This removes $H_{c}$ from the model and reduces
the gauge charge constraint to the simpler form $A_{r}={+}({-})1$.
The remaining $\mathbb{Z}_{2}$ gauge theory $H_{g}$ is referred
to as even(odd) \citep{Wegner1971,Moessner2001a}. It can also be
viewed as a toric code on the triangular ladder with a star energy
of $J_{S}={-}({+})\infty$.

A brief digression may be helpful to recap the $\mathbb{Z}_{2}$ gauge
theory on the square lattice. By duality, its even case is related
to the transverse field Ising model (TFIM) \citep{Wegner1971}, while
the odd case maps to the fully frustrated TFIM (FFTFIM) \citep{Moessner2001a,Moessner2001,Senthil2000}.
Extensive knowledge about both cases has been gathered \citep{Sachdev2018}.
Both undergo a deconfinement-confinement transition versus $h$, where
the low-$h$ phase -- the toric code descendant -- is topologically
ordered. In the odd case, frustration of the FFTFIM renders the quantum
phases significantly more complex, comprising additional hidden symmetries
and spontaneously broken translational invariance. More details can
be found in refs. \citep{Blankschtein1984,Huh2011,Wenzel2012}.

\begin{figure}[tb]
\centering{}\includegraphics[width=0.8\columnwidth]{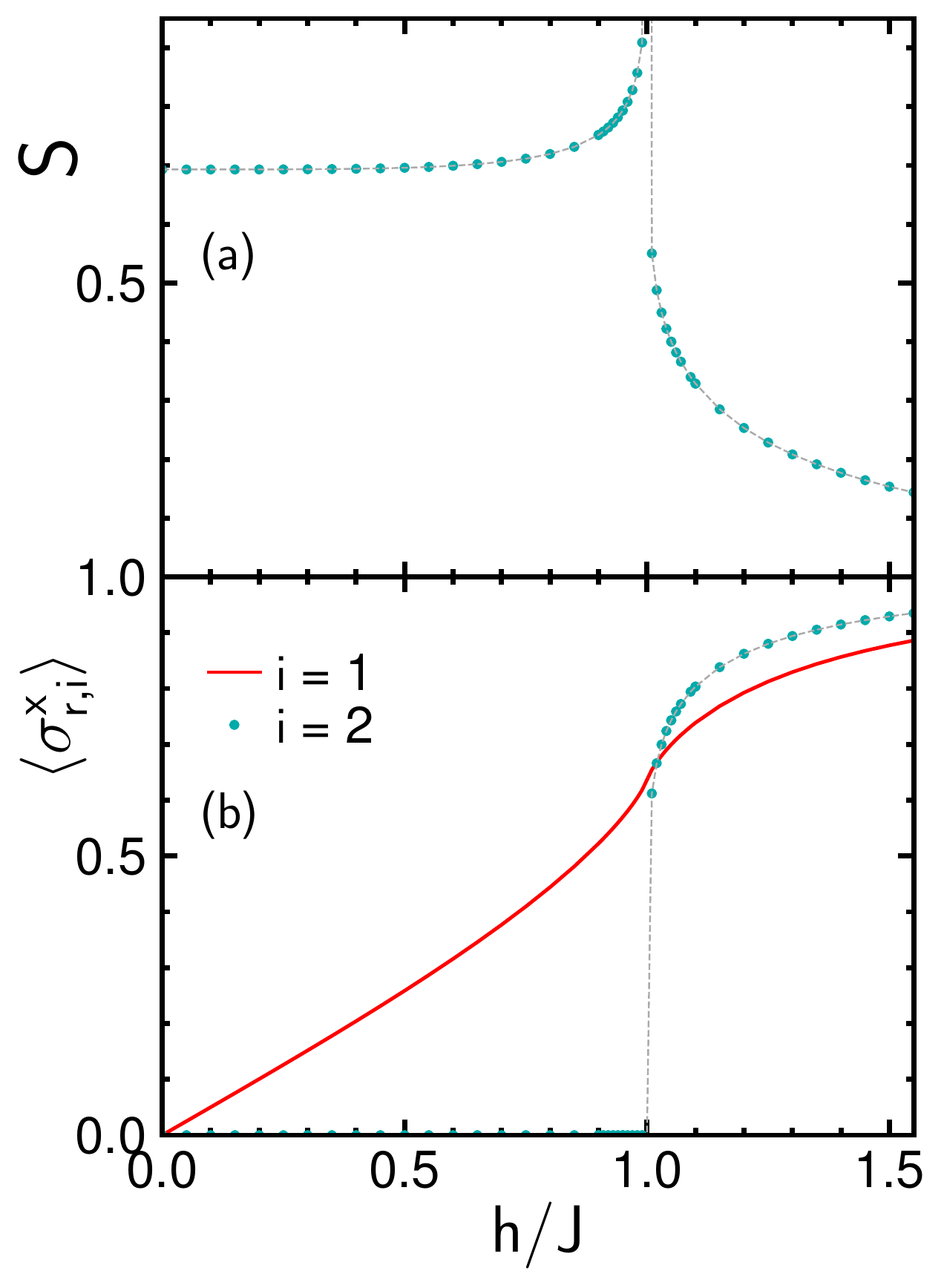}\caption{\label{fig:chainfield}(Color online) Chain-coupling: (a) Entanglement
entropy $S$ versus $h$ (turquoise dots). (b) Electric field expectation
values versus $h$ (red solid: $\langle\sigma_{r,1}^{x}\rangle$,
turquoise dots: $|\langle\sigma_{r,2}^{x}\rangle|$). Gray dashed
lines: guide to the eye. iDMRG, bond dimension $264$.}
\end{figure}

\begin{figure}[tb]
\centering{}\includegraphics[width=0.8\columnwidth]{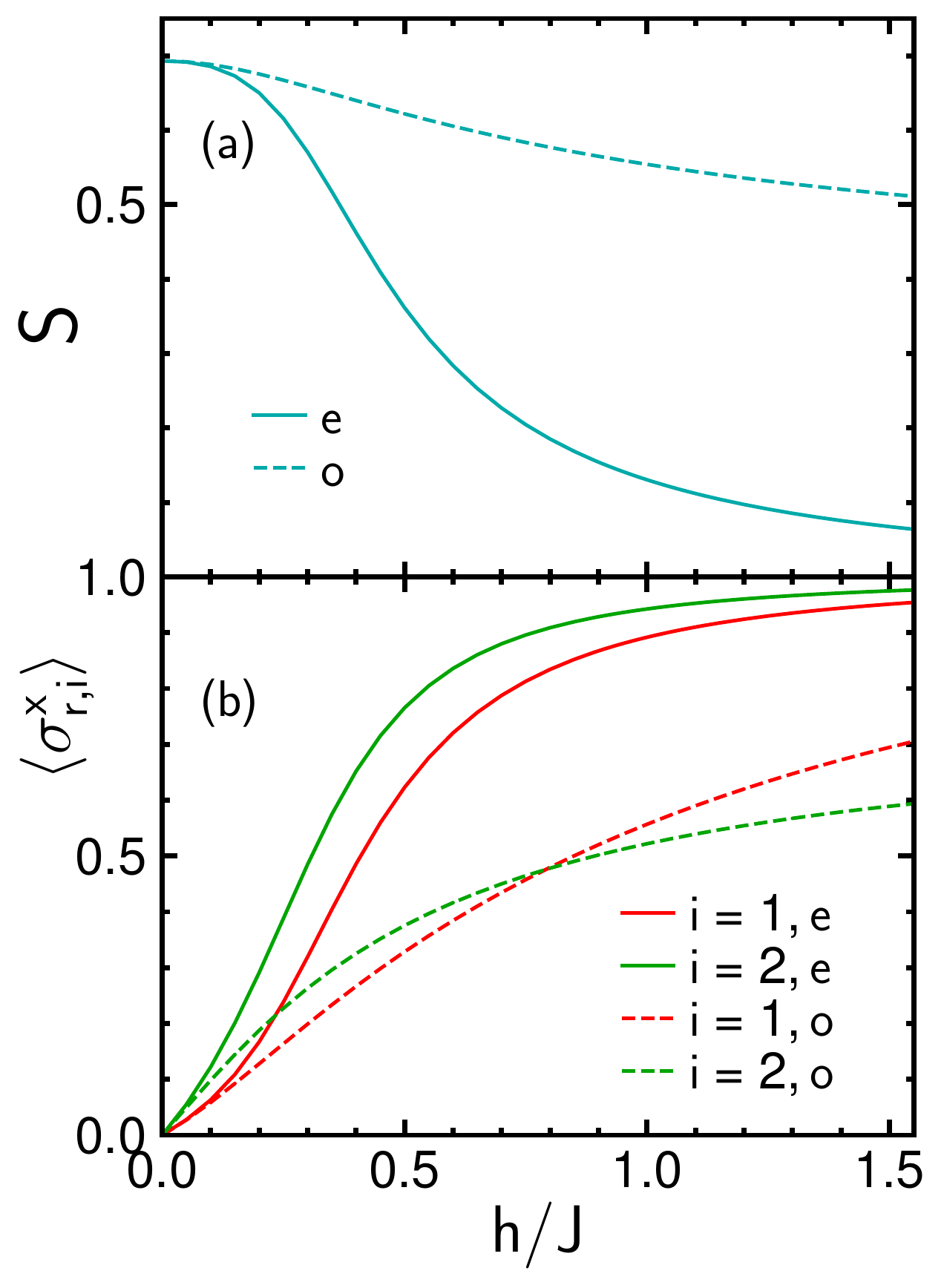}\caption{\label{fig:ladderfield}(Color online) Ladder-coupling for even (solid,
label $e$) and odd (dashed, label $o$) theory. (a) Entanglement
entropy $S$ versus $h$. (b) Electric field expectation values versus
$h$ (red: $\langle\sigma_{r,1}^{x}\rangle$, green: $\langle\sigma_{r,2}^{x}\rangle$).
iDMRG, bond dimension $264$.}
\end{figure}

As compared to the square lattice, the even and odd gauge theories
on the triangular ladder are different. First, for $h_{r,i}=0$, unitary
transformations $U=\prod_{b}\sigma_{b}^{z}$ can be formulated, using
selected subsets of links $b$ on the ladder, such that $U^{+}A_{r}U=-A_{r}$,
$\forall r$. E.g., $b$ can be chosen to comprise all odd rungs,
or each second segment of both legs. This implies that even and odd
gauge theories are identical for $h_{r,i}=0$. Second, since $U$
can be chosen to commute with $h\sum_{r}\sigma_{r,1}^{x}$, the even
and the odd gauge theories are also identical for finite chain-coupling.
Third, only for chain-coupling a critical behavior similar to the
one-dimensional TFIM can be expected versus $h$, since only the single
$h_{r,1}$-link exists between each nearest-neighbor pair of triangular
plaquettes along the linear direction of the ladder, while on the
legs $h_{r,2}=0$. For ladder-coupling, the dangling terms $h\sigma_{r,2}^{x}$
on the legs break the correspondence to the TFIM. Fourth, the preceding
unitary $U$ cannot be chosen to commute with all of $h\sum_{r,i=1,2}\sigma_{r,i}^{x}$.
Therefore, when transforming the odd gauge theory for ladder-coupling,
it will map to an even theory with a finite fraction of electric field
energies having a reversed sign. In turn, the expectation values $\langle\sigma_{r,i}^{x}\rangle$
versus $h$ will display a weaker increase with $h$ in the odd case
for ladder-coupling, as compared to the even one. Finally, the ladder
is quasi-1D and therefore, topological order with fourfold ground
state degeneracy, as for the 2D square-lattice toric code cannot be
claimed. Nevertheless, the ground state at $h_{r,i}=0$ is a twofold
degenerate loop-gas, the two states of which can be labeled by the
parity of $\sigma_{r,i}^{x}$ eigenvalues along any cut, comprising
one rung- and two leg-bonds.

Next, in Figs. \ref{fig:chainfield} and \ref{fig:ladderfield}, the
preceding is considered from a numerical point of view, using iDMRG
from the TeNPy library\citep{Hauschild2018} on Eq (\ref{eq:3}) with
$J=\pm$1. For the iDMRG an initial cell of $L=4$ $r$-sites, comprising
$4\times2$ spins, see Fig. \ref{fig:lat}, has been used, such as
to comprise a single star, both on the lower and the upper leg of
the ladder. Fig. \ref{fig:chainfield} refers to chain-coupling and
therefore applies to both, the even and the odd theory. Fig. \ref{fig:chainfield}(a)
shows the entanglement entropy. It displays the anticipated quantum
phase transition, similar to that of the TFIM, with a critical coupling
of $h_{c}/J=1$. The entropy at $h=0$ is $\ln(2)$. In Fig. \ref{fig:chainfield}(b),
the expectation values of the electric fields $\langle\sigma_{r,i}^{x}\rangle$
are depicted versus $h$. The electric field on the links connecting
the plaquettes, i.e., $\langle\sigma_{r,1}^{x}\rangle$, clearly shows
an increase in slope at $h_{c}$, translating into a peak in the susceptibility
$\chi^{x}(h)=\partial\langle\sigma_{r,1}^{x}\rangle/\partial h$ at
the critical point. This plot is very reminiscent of similar results
for the toric code on the square-lattice \citep{Trebst2007}. The
panel also shows the accompanying electric field on the legs, i.e.,
$\langle\sigma_{r,2}^{x}\rangle$. It is directionally degenerate,
i.e. Fig. \ref{fig:chainfield}(b) actually displays $|\langle\sigma_{r,2}^{x}\rangle|$.

Turning to ladder-coupling in Fig. \ref{fig:ladderfield}, one observes
no critical behavior. Both, the entanglement entropies in \ref{fig:ladderfield}(a)
as well as the expectation values of the electric fields in \ref{fig:ladderfield}(b),
are smooth functions of the coupling constant $h$. Both panels clearly
follow the previously made assertion of a different behavior of the
even versus the odd theory, with a weaker response of the odd theory
to $h$.

Summarizing this subsection, apart from the absence of 2D topological
order, the triangular ladder differs significantly from the square
lattice case regarding the distinction between even and odd phases,
and regarding the different action of electric chain- versus ladder-coupling.
The remainder of this work focuses on ladder-coupling.

\subsection{$h=0$~: Static gauge theory at finite fermion density\label{subsec:stagphi}}

The strategy to handle the static case has been set forth in refs.
\citep{Prosko2017,Borla2021} and is independent of the type of lattice.
The idea is to map the original gauge-dependent fermions $c_{r}^{(\dagger)}$
and hopping matrix elements $t_{i}\sigma_{r,i}^{z}$ onto new gauge-invariant
fermions $d_{r}^{(\dagger)}$ and hopping matrix elements $\gamma_{r,i}t_{i}$,
where $\gamma_{r,i}$ is a classical variable. This is achieved by
defining $d_{r}^{(\dagger)}$ via the non-local operator $d_{r}^{(\dagger)}=c_{r}^{(\dagger)}\prod_{b}^{\infty}\sigma_{b}^{z}$,
or equivalently $c_{r}^{(\dagger)}=d_{r}^{(\dagger)}\prod_{b}^{\infty}\sigma_{b}^{z}$,
where the product over $\sigma_{b}^{z}$ represents a semi-infinite
string, starting on any bond of the star centered at $r$, and extending
to infinity. 'Semi-infinite' implies that for each site $r'\neq r$
which the string passes through, it will share two of its bonds $b$
with the star of $G_{r'}$. The actual path of the string can be chosen
arbitrarily. Here, a path is used that extends right to the fermion
sites, along the corresponding legs. In any case, $G_{r}d_{r}^{(\dagger)}G_{r}=d_{r}^{(\dagger)}$,
i.e., the new fermions are indeed gauge-invariant, moreover $n_{r}=c_{r}^{\dagger}c_{r}^{\phantom{\dagger}}=d_{r}^{\dagger}d_{r}^{\phantom{\dagger}}$.

\begin{figure}
\centering{}\includegraphics[width=0.85\columnwidth]{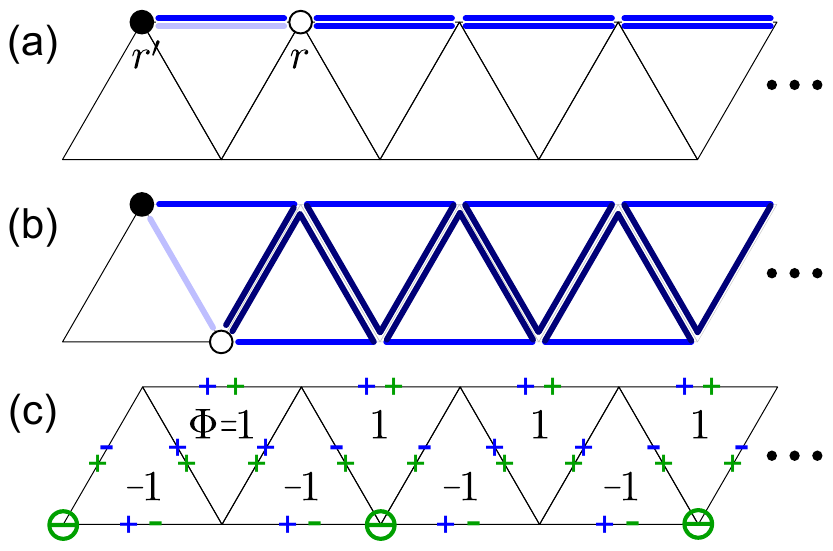}\caption{\label{fig:static}(Color online) Hopping processes (a) $t_{2}$ and
(b) $t_{1}$ on a link comprising one $\mathbb{Z}_{2}$ Peierls-factor
$\sigma_{r,i}^{z}$ (light blue) and gauge-dependent fermions $c_{r^{(\prime)}}^{(\dagger)}$:
Constructed from gauge-invariant fermions $d_{r^{(\prime)}}^{(\dagger)}$
(open (solid) black circle) attached to string $\prod_{b}^{\infty}\sigma_{b}^{z}$
(blue links). Dark blue links: Auxiliary pairs of $\sigma_{b}^{z}$
inserted to complete plaquettes. All pairs of $\sigma_{b}^{z}$ square
to unity. (c) Distribution of $\gamma_{r,i}$ in staggered flux state
for $J=\mu=0$ (blue signs). Green signs: $\gamma_{r,i}$ for identical
state with every $d_{r}^{(\dagger)}$ fermion at green $\ominus$-site
gauged to $-d_{r}^{(\dagger)}$.}
\end{figure}

The transformation of the kinetic energy is depicted in Fig. \ref{fig:static}.
While on the legs, the semi-infinite strings and the $\mathbb{Z}_{2}$
Peierls-factor square to $1$, on the rungs, they can be augmented
by a semi-infinite product of (conserved) plaquette operators $B_{r}\equiv\pm1$.
This turns the spinless fermion Hamiltonian into
\begin{equation}
H_{c}^{0}=-\sum_{r,i=1,2}(t_{i}\gamma_{r,i}d_{r+i}^{\dagger}d_{r}^{\phantom{\dagger}}+h.c.)-\mu\sum_{r}n_{r},\label{eq:6}
\end{equation}
with gauge-invariant fermions and the classical variables $\gamma_{r,1}=\pm1$
and $\gamma_{r,2}=1$.

Moreover, using the $\gamma_{r,i}$, and while the plaquettes $B_{r}$
from Eq. (\ref{eq:4}) certainly are quantum operators, their eigenvalues,
which remain conserved for $h_{r,i}=0$, can be expressed by the classical
fluxes $\Phi_{r}=\prod_{b\in P_{r}}\gamma_{b}$. In turn, finding
the ground state of the model Eqs. (\ref{eq:2},\ref{eq:3}) reduces
to minimizing the energy of
\begin{equation}
H_{c}^{0}-J\sum_{r}\Phi_{r}\,,\label{eq:7}
\end{equation}
with respect to the variables $\gamma_{r,i}$. Depending on the optimum
$\Phi_{r}$-pattern and the lattice structure, metals, semi-metals,
and insulators of the $d$-fermions may result.

\begin{figure}[tb]
\centering{}\includegraphics[width=0.8\columnwidth]{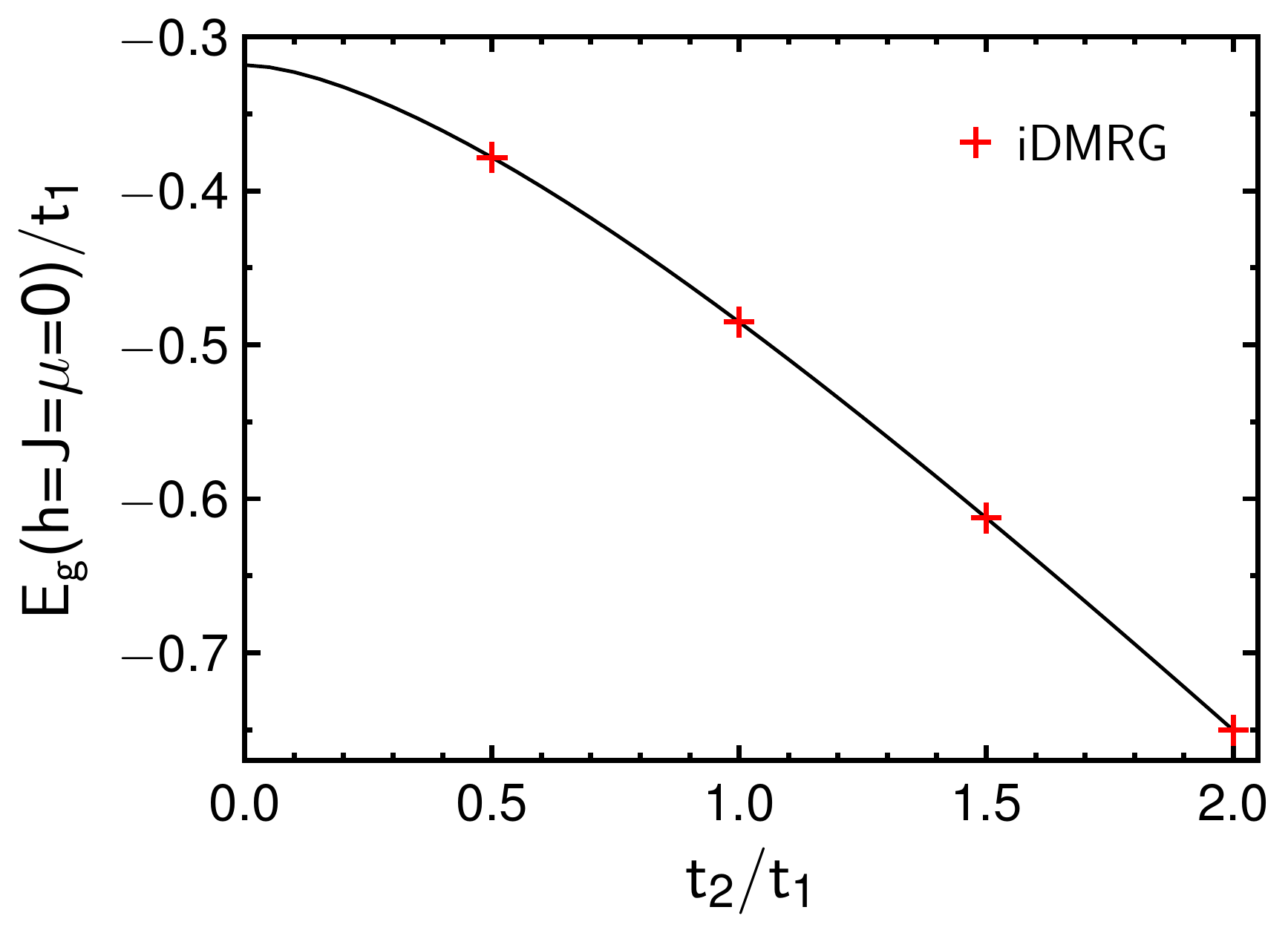}\caption{\label{fig:7}(Color online) Ground state energy of staggered flux
state of static gauge theory at half-filling versus $t_{2}/t_{1}$.
Solid: Analytic result Eq. (\ref{eq:8}). Red crosses: iDMRG.}
\end{figure}

On bipartite graphs, and for $J,\mu=0$, it has been proven in ref.
\citep{Lieb1994} that models of type (\ref{eq:7}) will acquire a
$\pi$-flux-phase ground state. The triangular ladder is a different
graph. Yet, it is straightforward to check that the state of lowest
energy for the model in that case is a \emph{staggered} flux state.
As can be read off from Fig. \ref{fig:static}(c), its spectrum can
also be obtained from free fermions hopping on the ladder, with all
identical signs on the rungs and a sign-flip between the upper and
lower leg. The dispersion in the latter gauge reads
\begin{equation}
\varepsilon_{k}^{\pm s}=\pm2\sqrt{t_{1}^{2}\cos(k)^{2}+t_{2}^{2}\cos(2k)^{2}}\,.\label{eq:8}
\end{equation}
The lattice constant is $2$ and $k\in[-\pi/2,\pi/2]$ is the Brillouin
zone (BZ). For any nonzero $t_{1}$ and $t_{2}$, this represents
a \emph{band insulator}. It features a gap of $\text{\ensuremath{\Delta}}=4|t_{2}|$
at $k=\pi/2$ if $t_{2}/t_{1}<1/2$, or $\Delta=|t_{1}|[8-(t_{1}/t_{2})^{2}]^{1/2}$
at $k=\pi/2-\arctan([16(t_{2}/t_{1})^{4}-1]^{1/2})/2$ if $t_{2}/t_{1}\ge1/2$.
The ground state energy per site of the spin-chain representation
is $E_{g}^{s}(h{,}J{,}\mu{=}0)=-\int_{0}^{\pi/2}\varepsilon_{k}^{+s}dk/(2\pi)$
which is $1/4$ of the ground state energy per unit cell of the fermion
model (\ref{eq:6}).

The spontaneous breaking of the symmetry between the sign of the hopping
integral on the two legs has a consequence for the local fermion density.
Namely, while $n_{r}$ is homogeneous on each individual leg and $n_{r}+n_{r+1}=1$
for $\mu=0$, at any finite ratio of $t_{2}/t_{1}$, the difference
$n_{r}-n_{r+1}$ is finite. I.e., there is a spontaneous symmetry
breaking of the fermion density between the legs. This can be understood
by realizing that, at half filling and for $t_{1}=0$, essentially
BZ-'center' ('boundary') states are occupied on the leg with $t_{2}\gamma_{r}<(>)\,0$.
Mixing these at finite $t_{1}$ lifts their balance of local densities.
An elementary calculation yields
\begin{equation}
n_{r}^{\pm}-\frac{1}{2}=\pm\frac{2}{\pi}\int_{0}^{\pi/2}t_{2}\cos(2k)/\varepsilon_{k}^{+s}dk\,.\label{eq:25}
\end{equation}
Since either for $t_{1}=0$, or for $t_{2}=0$, one has $n_{r}=1/2$,
$\forall r$, the right-hand side of Eq. (\ref{eq:25}) has an extremum
at some intermediate $t_{2}/t_{1}|_{\mathrm{ex}}$. One finds approximately
$t_{2}/t_{1}|_{\mathrm{ex}}\simeq0.35355$, with $n_{r}^{+}|_{\mathrm{ex}}-1/2\simeq0.07735$.

\begin{figure}
\centering{}\includegraphics[width=0.95\columnwidth]{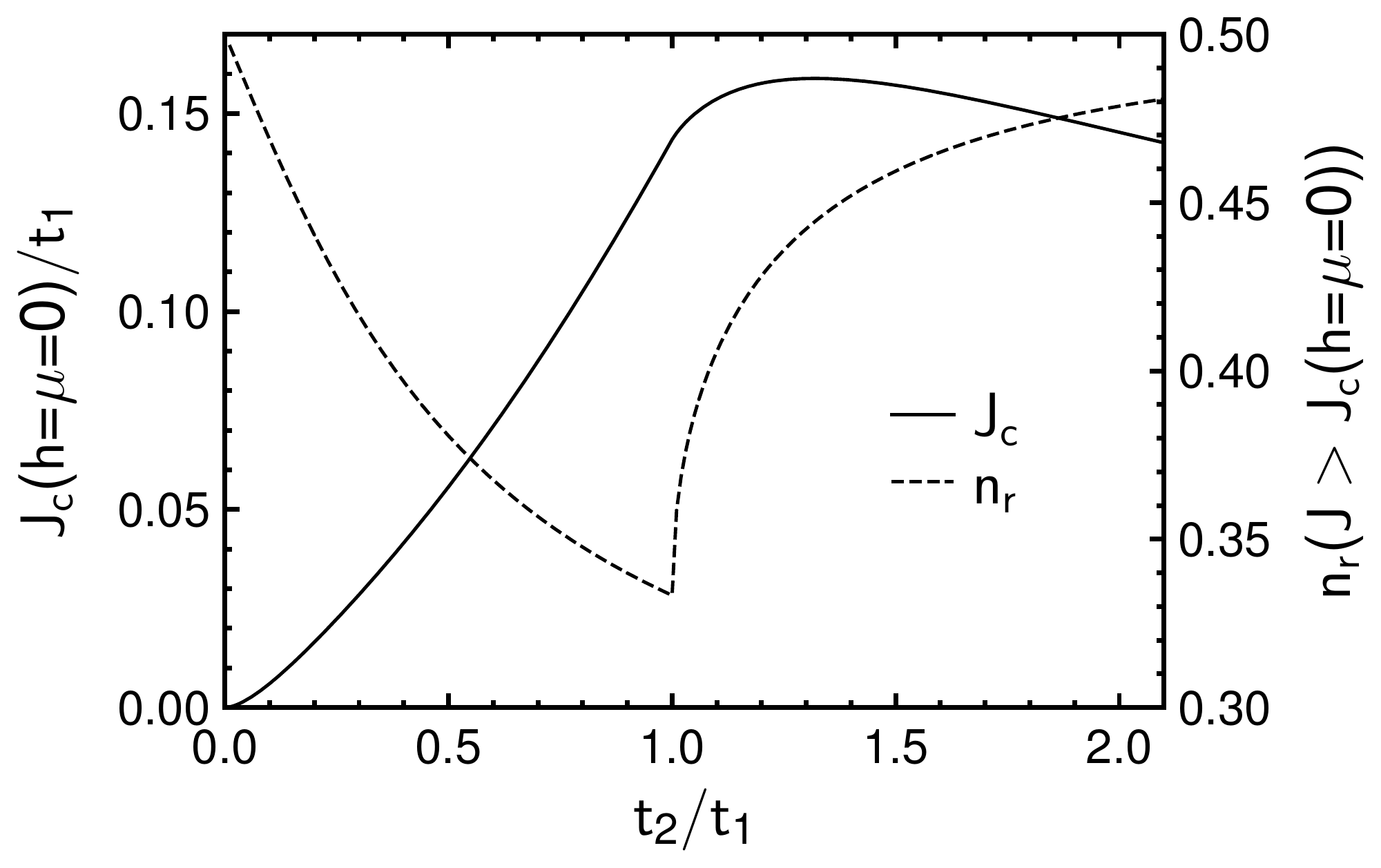}\caption{\label{fig:7a}(Solid) First order quantum critical line in the $(J/t_{1,}t_{2}/t_{1})$-plane
between the low-$J$ staggered flux band-insulator and the large-$J$
uniform flux metal at $\mu=0$. (Dashed) On-site fermion number $n_{r}$
in the uniform flux phase at $\mu=0$ versus $t_{2}/t_{1}$.}
\end{figure}

In Fig. \ref{fig:7} the ground state energy obtained from both, the
analytic result for $\varepsilon_{k}^{\pm s}$, and from iDMRG for
a selected set of points is shown versus $t_{2}/t_{1}$. These results
obviously agree very well. It should be noted that in performing the
iDMRG analysis, it has also been checked that indeed, the flux expectation
value is staggered along the ladder, and moreover, that using a small
pinning potential one can switch between its two degenerate staggering
sequences. Without explicit display and needless to say, the local
fermion density obtained from the iDMRG is indeed equal to the analytic
result Eq. (\ref{eq:25}).

For $J\gg t_{1,2}$ and from Eq. (\ref{eq:7}), a uniform flux state
with $\Phi_{r}=1$, $\forall r$ is favored. Here, the size of the
unit cell is 1 and $k\in[-\pi,\pi]$ is the BZ. However, to ease comparison
with the staggered state, the unit cell is enlarged to size 2, keeping
a BZ of $k\in[-\pi/2,\pi/2]$ and zone-fold the fermion dispersion
by $\pi$ onto two bands, i.e. $\varepsilon_{k}^{\pm u}=\pm2t_{1}\cos(k)-2t_{2}\cos(2k)$,
see \citep{Note2}. For any filling $0<n_{r}<1$, this represents
a \emph{simple metal}.

In contrast to the staggered flux state, Eq. (\ref{eq:8}), $\varepsilon_{k}^{\pm u}$
is \emph{not} particle-hole symmetric. In turn, the transition from
the staggered to the uniform flux state differs for a micro-canonical
versus a canonical setting. Here, the latter is considered and $\mu=0$
is used. This implies that at the transition the fermion number jumps
discontinuously. While the Fermi-points for $\varepsilon_{k}^{\pm u}$
and the uniform ground state energy $E_{g}^{u}$ at $\mu=0$ can be
determined analytically, $E_{g}^{s}$ requires numerical integration.
The transition line obtained from comparing Eq. (\ref{eq:7}) for
the two cases is depicted in Fig. \ref{fig:7a}. The singular behavior
at $t_{2}/t_{1}=1$ is related to the bottom of the band $\varepsilon_{k}^{+u}$
crossing $\mu$, i.e., zero. The asymptotic behavior of $J_{c}$ follows
from $\varepsilon_{k}^{-s}\rightarrow\varepsilon_{k}^{-u}$ for $t_{2}/t_{1}\rightarrow0$,
while for $t_{2}/t_{1}\rightarrow\infty$ the sum of energies from
$\varepsilon_{k}^{\pm u}$ approaches that from $\varepsilon_{k}^{-s}$.

Concluding this section, several points are mentioned on the side.
First, all of the preceding obviously depends decisively on the lattice
structure and, therefore, is different from the square lattice case
of ref. \citep{Borla2021}. In the latter, the QPT versus $J$ occurs
between a Dirac and a conventional metal. Second, for this work it
remains an open question if the staggered to uniform transition would
allow for additional intermediate phases with more complicated flux
patterns. This could be clarified by classical Monte-Carlo analysis.
Finally, the microcanonical case and also the dependence on general
filling fractions remain to be studied.

\begin{figure}[tb]
\centering{}\includegraphics[width=1\columnwidth]{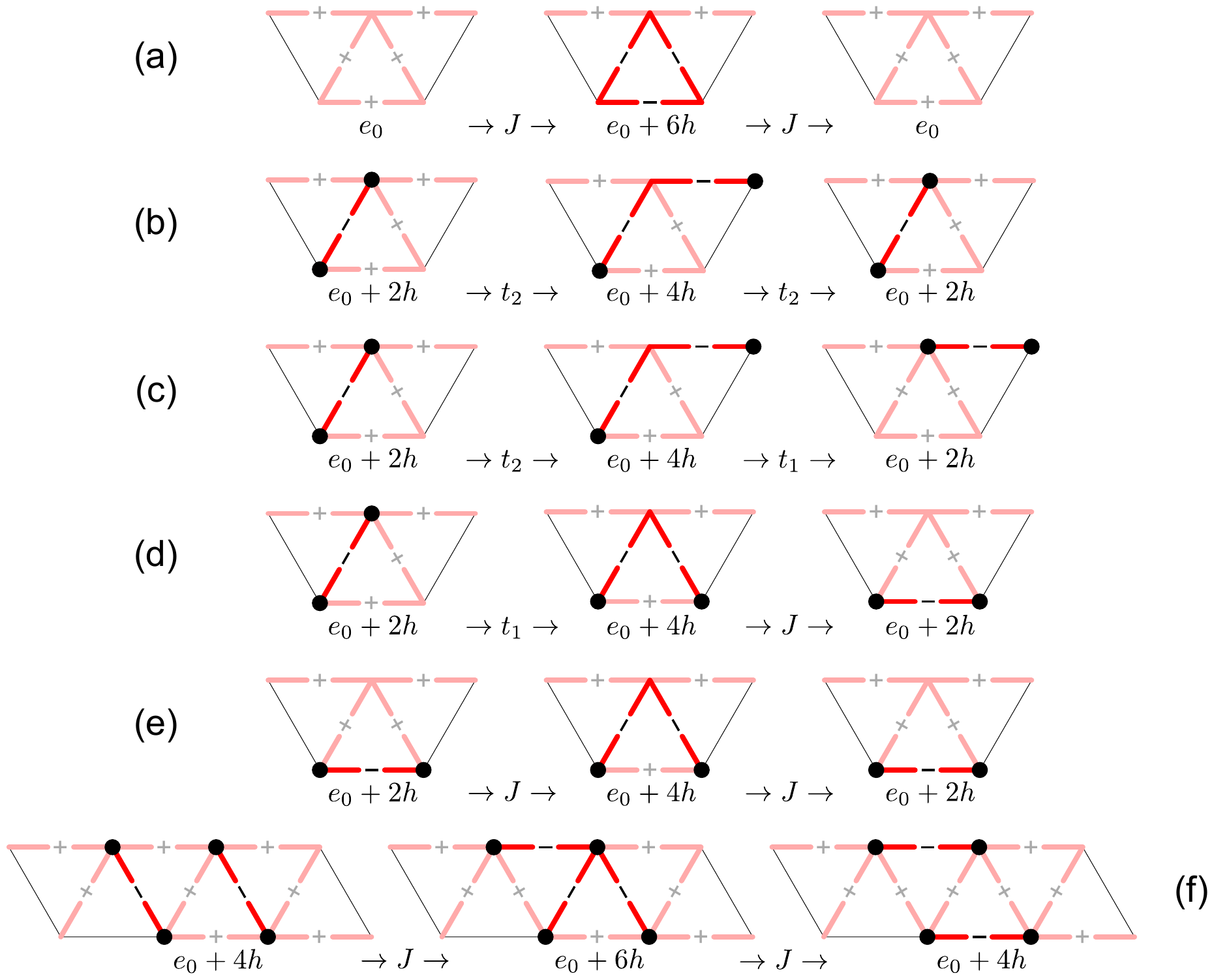}\caption{\label{fig:8}(Color online) Typical processes to $O(t_{1,2}^{2}/h,$
$Jt_{1,2}/h,J^{2}/h)$ in the large-$h$ limit for $0,2,$ and $4$
fermions. (a) ground state renormalization / vacuum fluctuations,
(b) single dimer dressing, (c,d) single dimer hopping, (e) single
dimer polarization, (f) two dimer resonance. For better visibility,
$\sigma^{x}=-1$ links, i.e., with increased string energy $2h$ shown
with higher contrast.}
\end{figure}

\subsection{$h\gg|t_{1,2}|,|J|,|\mu|$: Strong confinement \label{subsec:largeh}}

If the electric coupling is the largest energy scale, the spectrum
can be understood qualitatively by treating $t_{1,2}$ and $J$ perturbatively,
taking a microcanonical point of view, labeling the states by $|\nu,N\rangle$,
with $N$ being the total fermion number $\sum_{r}n_{r}|\nu,N\rangle=N|\nu,N\rangle$.
For the remainder of this subsection, ladder-coupling, i.e. $h_{r,1}=h_{r,2}=h$
is implied. The ground state is from the sector $|\nu,0\rangle$ and
for $J=0$ it has $\sigma_{r,i}^{x}=+1$ on all bonds with an energy
of $e_{0}/L=-2h$. The latter accounts for two links per unit cell
of Eq. (\ref{eq:6a}) on the green chain in Fig. \ref{fig:lat}. For
$J\neq0$, the plaquettes will lower the ground state energy to $O(J^{2}/h)$,
see Fig. \ref{fig:8}(a). The ground state is separated from all other
zero-fermion states by energies of at least $O(6h)$, resulting from
the application of odd numbers of plaquettes.

Within the aforementioned gap of $O(6h)$. Two types of states arise
with fermions present. These are two- and four-fermion states, $\{|\nu,2\rangle\}$
and $\{|\nu,4\rangle\}$, respectively. In both of these sectors,
and for $J=t_{1,2}=0$, the ground state minimizes the electric string
length. I.e., the fermions pair into \emph{dimers} on nearest neighbor
bonds with energies of $e_{2(4)}-e_{0}=2h\,(4h)$. Speaking differently,
this is a strongly \emph{confined} phase.

To enumerate the possible single dimer processes, recall from Eq.
(\ref{eq:2}) that hopping fermions will always flip the string state
on the bond, with the string tracing the hopping path. In turn, the
final state of a hop does not necessarily comprise the lowest electric
energy state. E.g., hopping one fermion of a dimer from one corner
of a triangle to another, terminates in an excited state of $\tilde{e}_{2}-e_{0}=4h$
with a string length of $2$. In turn, there is no resonance of dimers
on triangles at $O(t_{1,2})$.

\begin{figure}[tb]
\centering{}\includegraphics[width=0.95\columnwidth]{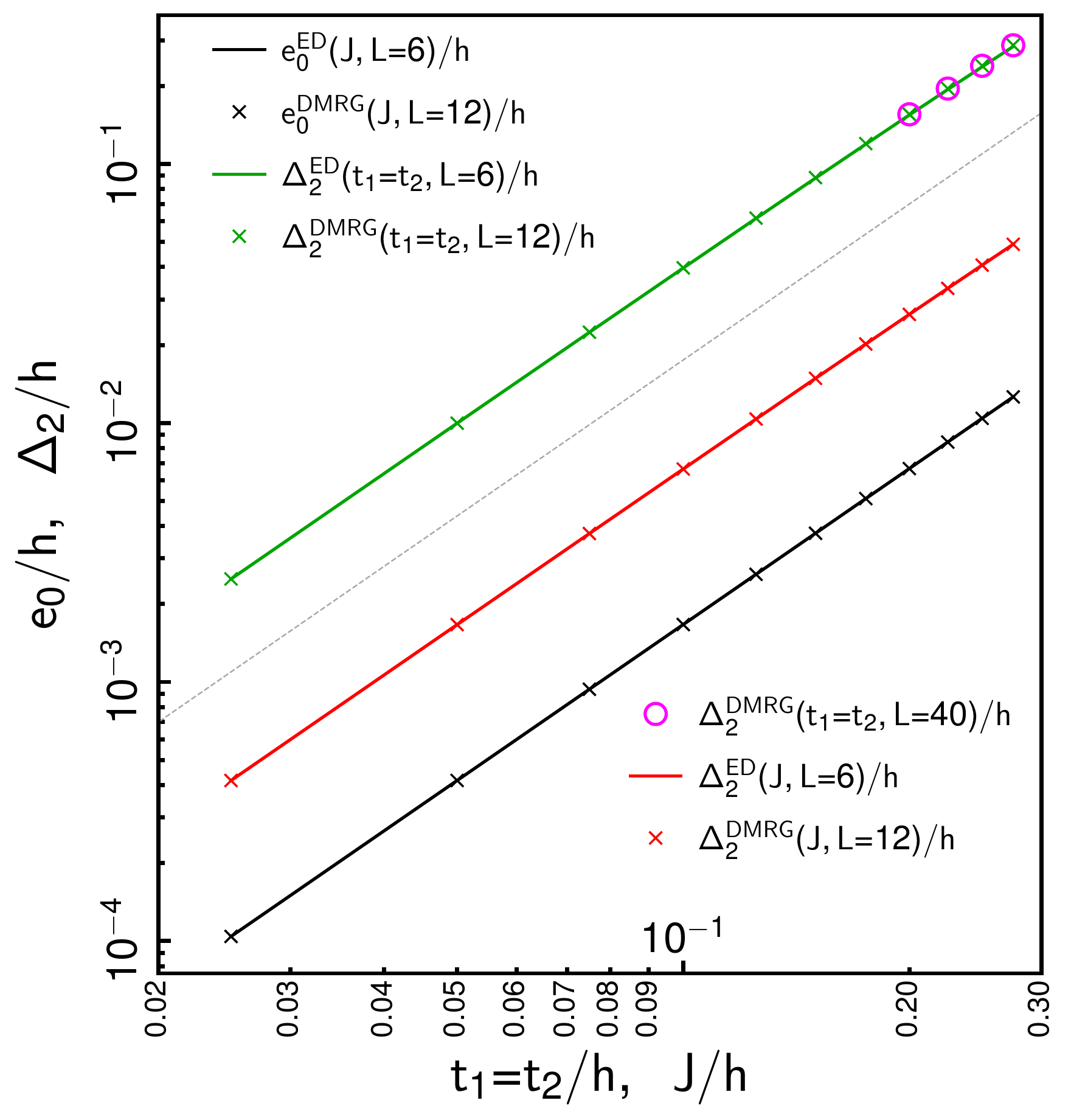}\caption{\label{fig:9}(Color online) Log-log plot of ground- and first excited-state
energies at large $h$. Black: Ground state energy $e_{0}/h=E_{0}/(Lh)$,
per site, i.e., per two spins, versus $J$. Green: Two-fermion excitation
energy $\Delta_{2}/h$ versus $t_{1}=t_{2}$. Red: Two-fermion excitation
energy $\Delta_{2}/h$ versus $J$. Solid curves: ED on $L=6$ sites,
i.e., $12$ spins. Crosses(open circles): DMRG on $L=12$($40$) sites,
i.e., $24$($80$) spins, with bond dimension $40$($100$). Thin
dashed gray: $y\propto x^{2}$ for reference.}
\end{figure}

At higher orders, and for $J=0$, but $t_{1,2}\neq0$, single dimers
can lower their bare on-bond energy of $2h$ with a \emph{polarization
cloud}, as in Fig. \ref{fig:8}(b), and they can hop, as in Fig. \ref{fig:8}(c),
both at $O(t_{1,2}^{2}/h)$. With both, $J\neq0$ and $t_{1,2}\neq0$,
mixed hopping processes at $O(Jt_{1,2}/h)$ become available, see
Fig. \ref{fig:8}(d). Finally, for $J\neq0$, but $t_{1,2}=0$, single
dimers can again lower their bare on-bond energy by polarization processes
of type of Fig. \ref{fig:8}(e). This does indeed lower the energy,
despite the vacuum fluctuations of Fig. \ref{fig:8}(a), because for
the latter, the intermediate state energy is larger by $2h$. As dimer
hopping does not occur for $t_{1,2}=0$, the gap is degenerate at
least to $O(L)$ in that case.

Turning to two dimers, i.e., four-fermion states, they experience
two types of irreducible interactions, beyond the single dimer dynamics.
First, for $t_{1,2}\neq0$, the lowering of a single dimer energy
by polarization processes of type Fig. \ref{fig:8}(b) are Pauli-blocked,
if another dimer occupies sites of the intermediate state. Therefore,
a short-range repulsion of $O(t_{1,2}^{2}/h)$ exists between dimers.
Second, and for $J\neq0$, nearby pairs of dimers can lower their
energy by a \emph{resonance} move, as in Fig. \ref{fig:8}(f). I.e.,
there exists a short range attraction of $O(J^{2}/h)$.

To summarize, at $t_{1,2}/h,J/h\ll1$, and for low fermion density,
the ladder hosts a gas of fermionic dimers of energies $2h$, which
hop and interact on (next-)nearest links on a scale of $O[(t_{1,2}^{2},Jt_{1,2},J^{2})/h]$.
Since in this limit the excitation gaps are large, all of the aforementioned
can be checked by numerical analysis on very small systems, since
finite size effects can be made negligible. In Fig. \ref{fig:9},
several energies are shown in this limit from exact diagonalization
(ED) for $L=6$, as well as from DMRG for $L=12$, and $40$ fermion
sites, i.e. for 12, 24, and 80 spins. Indeed these results are practically
independent of $L$ and are perfectly consistent with the quadratic
scaling versus $t_{1,2}$ and $J$.

\begin{figure}
\centering{}\includegraphics[width=0.9\columnwidth]{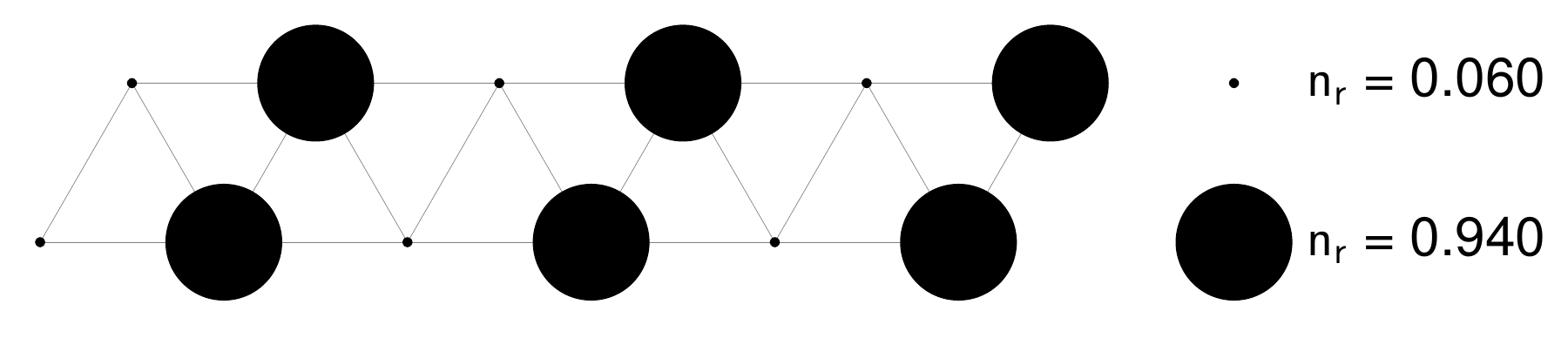}\caption{\label{fig:11}Dimer density wave at $t_{2}=1$ and $t_{1}=0.5$,
$h=4$, $\mu=4$, $J=0$. Size of solid black dots proportional to
fermion density $n_{r}$. Results are identical for iDMRG and DMRG
with $L=4$ and $L=100$, and at bond dimension $100$.}
\end{figure}

For finite fermion density at $t_{1,2}/h,J/h\ll1$, and with $t_{1,2}$
and $J$ both non-zero, the consequences of simultaneous dimer repulsion
and attraction are unclear at present. However, switching off attraction,
by setting $J=0$, and because of the off-site nature of the repulsion,
it is conceivable that dimer density waves (DDW) can form at suitable
fillings. This is confirmed by iDMRG calculations as depicted in Fig.
\ref{fig:11}, selecting a representative ratio of $t_{2}/t_{1}$,
at $n=\sum_{r}n_{r}/L=1$. Such DDWs may in addition be incompressible
(iDDW), i.e., $\partial n/\partial\mu=0$, implying a fermion particle
number gap $\Delta_{n}$. For the particular parameters used in Fig.
\ref{fig:11}, an iDMRG scan of $\mu$ indeed returns a gap of $\Delta_{n}/t_{2}\simeq0.29\pm0.02$
for $\mu/t_{2}\in[3.82,4.1]\pm0.01$. The error is rather large, since
iDMRG convergence at the gap edges turns out to be poor. It is likely
that iDDWs are a feature of the SFIGT for extended parameter ranges
at large $h/t_{1,2}\gg$1. A systematic search for them, scanning
$t_{1,2}/h$ and $n$, as well as an analysis of the scaling of their
gaps $\Delta_{n}$ with $t_{1,2}$, is beyond the scope of this work.

To close this subsection it should be emphasized again that also for
large $h$ the physics of the SFIGT described here strongly depends
on the lattice structure. Specifically, on the square lattice, the
confined dimers of the large-$h$ limit experience an attraction by
a resonance processes, occurring already at $O(J)$ \citep{Borla2021}.
In turn, one may speculate that the tendency for phase separation
of dimers in the confined phase is much less pronounced on the triangular
ladder than on the square lattice. This may also impact questions
of dimer BEC in that region.

\begin{figure}
\centering{}\includegraphics[width=0.95\columnwidth]{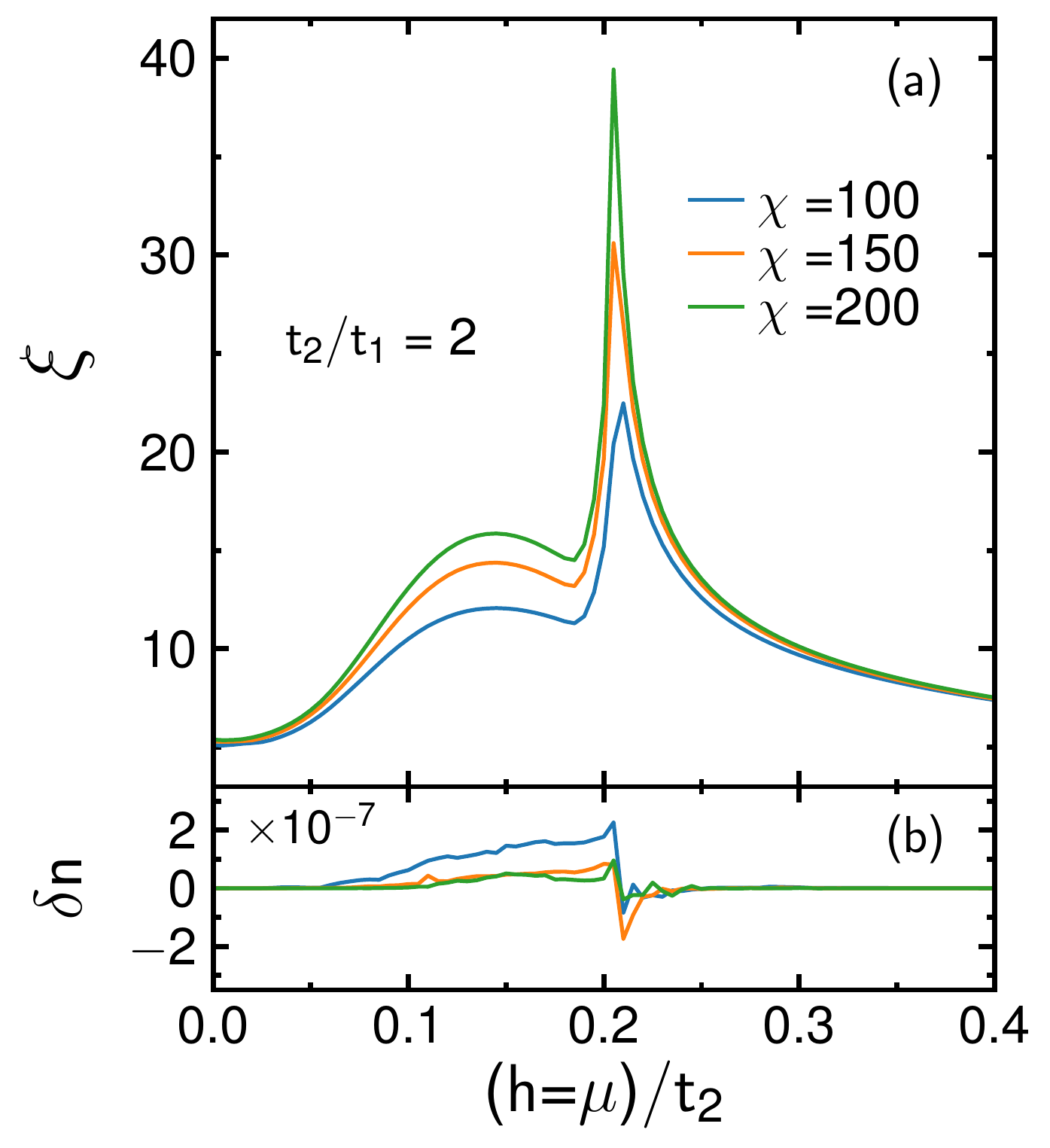}\caption{\label{fig:12}(Color online) (a) Correlation length $\xi$ versus
$h$ at half filling for increasing bond dimension $\chi=100\dots200$
(blue ... green). (b) Deviation from half filling for parameters identical
to panel (a). }
\end{figure}

\subsection{Staggered-flux insulator to iDDW transition\label{subsec:stagddwtrans}}

At $h,J=0$ the SFIGT at half filling, i.e. for $\mu=0$, is a band
insulator in the deconfined phase with a broken translational invariance
of the flux. For $h\gg t_{1,2}$ and at $J=0$, the iDDWs occurring
at half filling are correlation induced insulators in the confined
phase with no apparent flux order. A priori it is unclear if the deconfinement-confinement
transition, the iDDW formation, as well as the flux ordering occur
in a single or in multiple transitions. Similar questions are of great
interest on the square lattice for spinful \citep{Gazit2018}, as
well as for spinless fermions \citep{Borla2021}. In the former case
the transition to confinement comprises AFM ordering in addition and
leads to predictions of an emergent $SO(5)$ symmetry with
valence-bond states at criticality \citep{Gazit2018}.

Here, and following the idea of ref. \citep{Borla2021}, the correlation
length of the matrix product state (MPS) is considered versus $h$
in order to uncover quantum phase transitions. Luckily, keeping the
fermion number at $n=1$ while scanning $h$ can be achieved by setting
$\mu\simeq h$. In the two limiting cases, this follows by construction.
I.e., for $h=0$ and $t_{1,2}\neq0$, $\mu=0$ resides in the band
gap, while for $h/t_{1,2}\rightarrow\infty$ the dimer binding energy
of $2h$ in conjunction with the compressibility gap of the iDDW ensures
half filling. For intermediate $\mu$ the situation is not clear a
priori.

Fig. \ref{fig:12}(a) shows the correlation length $\xi(h)$, obtained
from iDMRG. From the behavior of $\xi$ versus bond dimension cut-off
$\chi$, it is clear that the system features only a single transition
at $h/t_{2}\approx0.2$ for $t_{2}/t_{1}=2$. Scanning this transition
with $t_{2}/t_{1}$ is left to future work. In addition, there is
a 'hump' at somewhat lower $h$ which may signal a crossover-behavior.
This is absent in previous studies of the SFIGT on the square lattice
\citep{Borla2021}. The origin of the hump is unclear at present,
however, it is worth mentioning that the relative height of the hump
can be varied by the ratio of $t_{2}/t_{1}$. Finally, Fig. \ref{fig:12}(b)
evidences a posteriori that $\delta n=(\sum_{r=1,2;i=1,2}n_{r,i}/L)-1/2$,
i.e., the deviation from half filling for $\mu=h$ and taking into
account the increase of the unit cell in the iDDW, remains zero up
to numerical errors over all of the relevant $h$-range.

\subsection{Finite-$J$ quantum phases in the $\mu$-$h$ plane \label{subsec:scan}}

\begin{figure}
\begin{centering}
\noindent\begin{minipage}[t]{1\columnwidth}%
\begin{center}
\includegraphics[width=1\columnwidth]{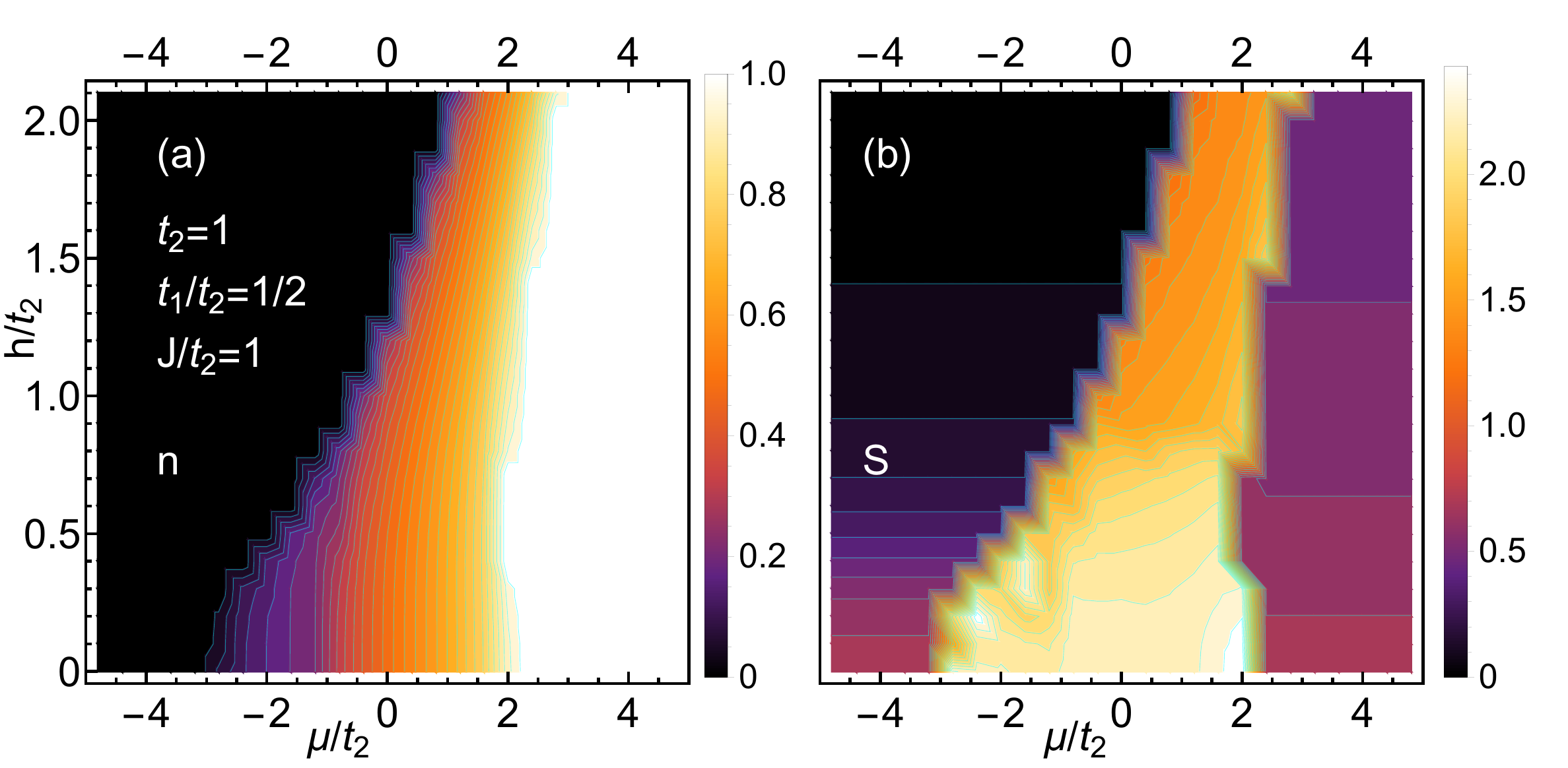}
\par\end{center}%
\end{minipage}
\par\end{centering}
\begin{centering}
\noindent\begin{minipage}[t]{1\columnwidth}%
\begin{center}
\includegraphics[width=0.61\columnwidth]{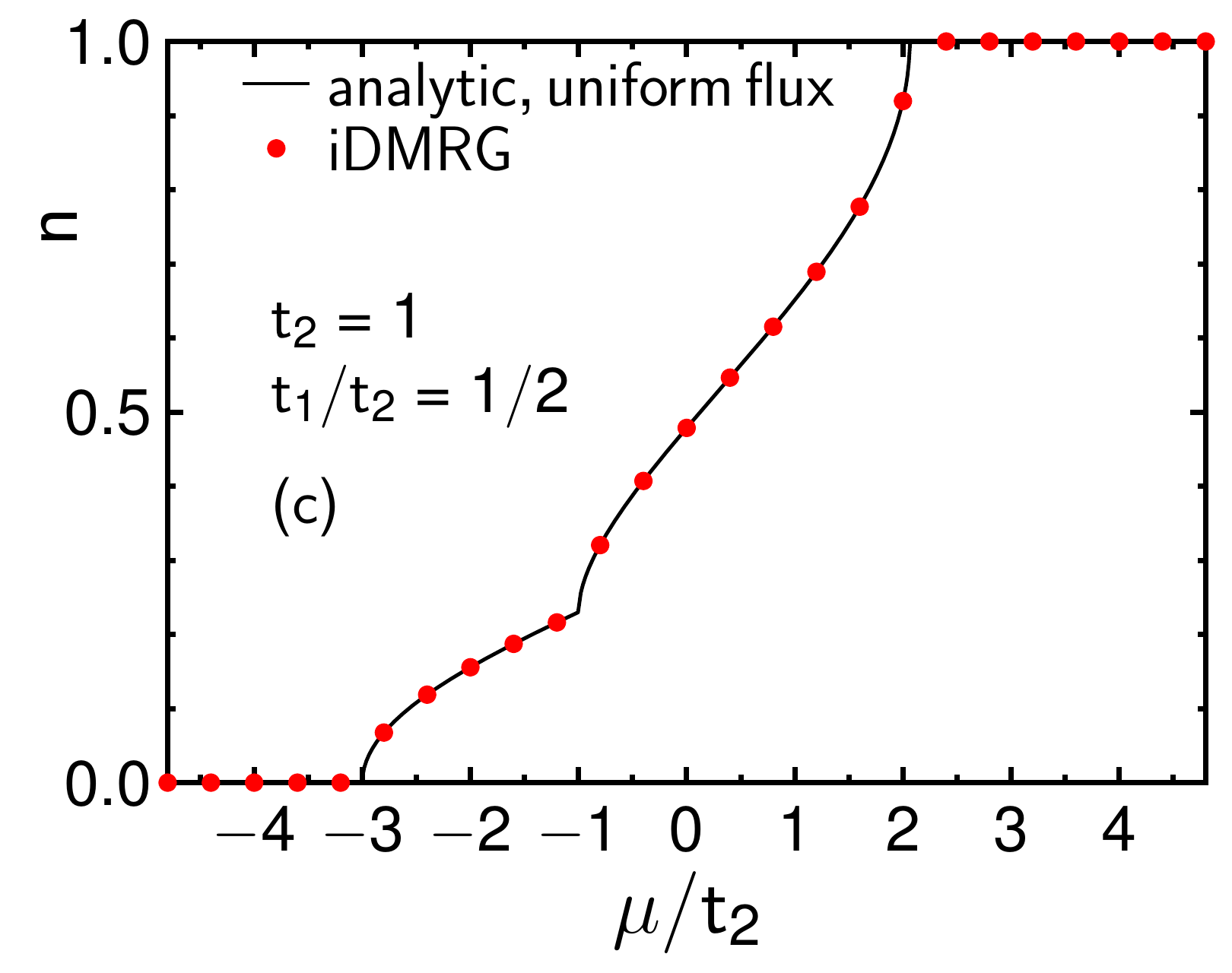}\includegraphics[width=0.39\columnwidth]{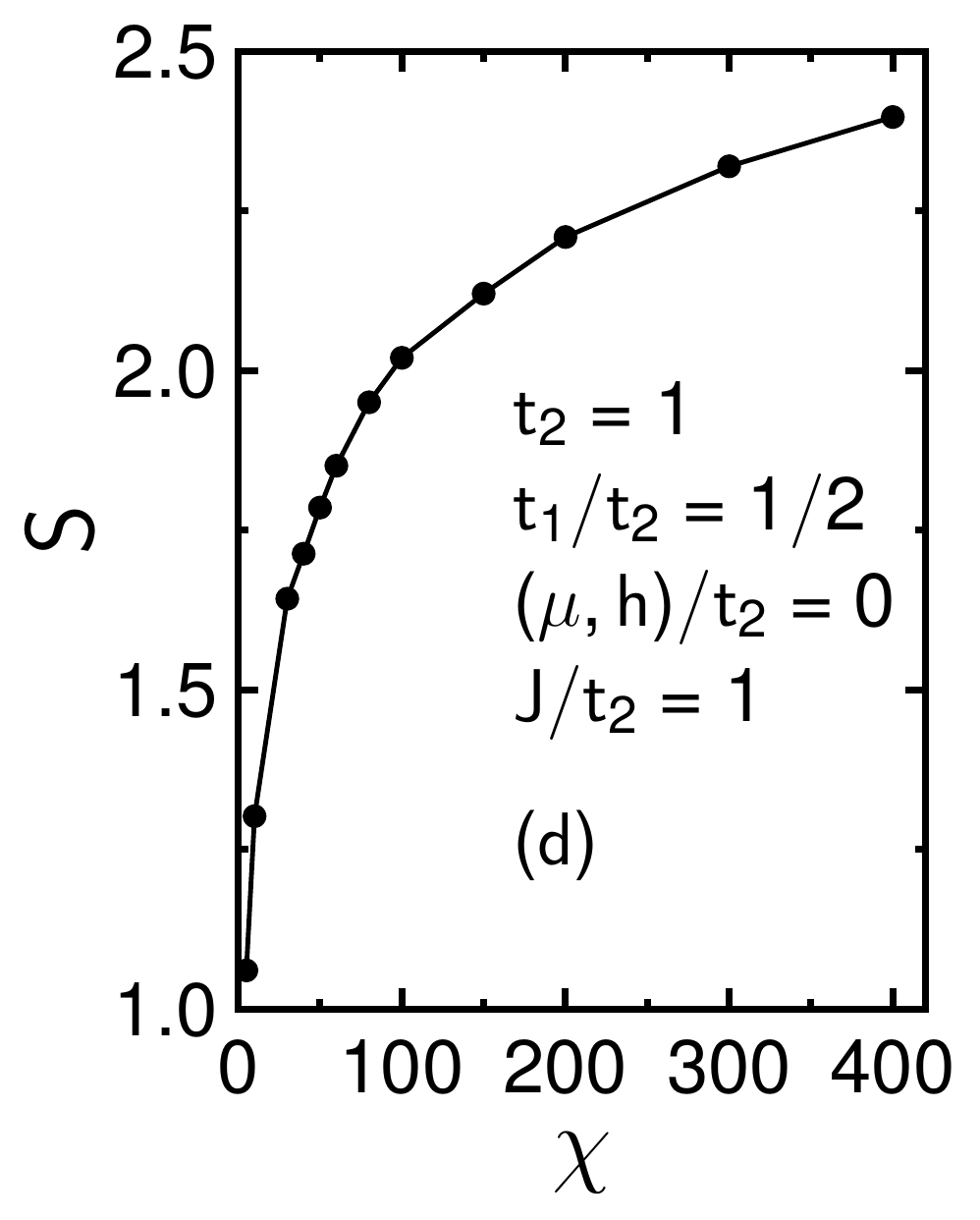}
\par\end{center}%
\end{minipage}
\par\end{centering}
\centering{}\caption{\label{fig13}(Color online) Contour plots of (a) fermion density
$n$ and (b) entanglement entropy $S$, in $\mu$-$h$ plane at finite
$J$, from iDMRG for $L=4$ at bond dimension $\chi=200$. (c) Comparing
analytic density (solid black line) in uniform flux phase with iDMRG
(red dots) on cut at $h=0$ from panel (a). (d) Entanglement entropy
at ($\mu,h$)-origin, in partially filled region versus bond dimension.}
\end{figure}

In this subsection, a coarse-grained overview is given over the quantum
phases versus filling and electric coupling at finite $J$ and $t_{1,2}$.
Ladder-coupling, i.e., $h_{r,1(2)}=h$ is used. Fig. \ref{fig13}(a)
and (b) display contours of the density $n$ and the entropy $S$,
respectively, in the $(\mu,h)$-plane, with a grid spacing of $(0.4,0.1)$.
Several comments are in order. First, in both panels, the three regions:
pure even, partially filled, and pure odd gauge theory can be distinguished
clearly from left to right. Second, the fermion band-width, which
can be read off from the region of partial filling, shrinks with increasing
electric coupling strength. I.e., there is a correlation induced mass
enhancement due to the confining interaction. Third, as $h$ increases,
the chemical potential for half filling starts to lean towards the
relation $\mu\simeq h$, signaling the dimer confinement energy. This
can be seen quite clearly for $h/t_{2}\simeq2$, at the upper edge
of Fig. \ref{fig13}(a), where for $n\simeq0.5$ one has to chose
$\mu/t_{2}\simeq2$. This relates directly to the choice of the chemical
potential used in Subsec. \ref{subsec:stagddwtrans}. Fourth, since
for $t_{2}/t_{1}=2$, $J=1$ is larger than $J_{c}$ for the transition
into the uniform flux state, see Fig. \ref{fig:7a}, the density versus
$\mu$ along a cut at $h=0$ in Fig. \ref{fig13}(a) can be obtained
from the analytic expression of the free fermion dispersion $\varepsilon_{k}^{\pm u}$
from Subsec. \ref{subsec:stagphi}. In Fig. \ref{fig13}(c) the latter
is compared to the iDMRG result from panel (a). The agreement is reassuring.

It is conceivable that similar to the case of $J=0$, also for $J>0$,
and for sufficiently large $h$, correlated iDDWs or related Mott-states
will form at suitable filling fractions. However, the density variations
observed on all $L=4$ central sites of the iDMRG are only small for
the parameters used in Fig. \ref{fig13}(a), which therefore displays
the site-averaged density. Nevertheless, the figure does not imply
only band-narrowing versus $h$ and does not rule out that analysis
with much higher resolution in $\mu,h$ would reveal incompressible
regions. Searching for such is clearly beyond the present study.

Turning to the entanglement entropy in Fig. \ref{fig13}(b), the crossover
between the deconfined and confined regions, exactly as discussed
for the two limiting cases of $\mu\rightarrow\pm\infty$ in Subsec.
\ref{subsec::Pure} and in Fig. \ref{fig:ladderfield}, can now be
seen to extend up to the lower and upper band-edges. Furthermore,
Fig. \ref{fig13}(b) also extends the 'greater sensitivity' of $S$
to the electric coupling in the even region as compared to the odd
one up to the band edges. While starting with $S(h=0)=\ln(2)$ both,
below and above the band edge, the fall-off of $S$ with $h$ above
the band edge is rather slow. In the partially filled region, the
interpretation of $S$ is less informative. First, the kinetic energy
in the effective chain model Eq. (\ref{eq:7a}) comprises two non-equivalent
bonds per unit cell due to $t_{1,2}$. Therefore, while the pure gauge
theories are insensitive to that, $S$ in the partially filled region
slightly differs on these two bonds. For simplicity, Fig. \ref{fig13}
displays a corresponding average of $S$. Second, at $h=0$, the uniform
flux phase is a gapless quasi-1D free spinless-fermion gas, which
likely is stable up to some finite $h$. In this region the entanglement
entropy is expected to scale logarithmically with system size \citep{Latorre2004},
being infinite in the thermodynamic limit. For iDMRG this implies
that $S$ will grow without bounds with the bond dimension. An example
of this is shown in Fig. \ref{fig13}(d) at $\mu,h=0$. Finally, if
dimer Mott-states exist at sufficiently large $h$, they could render
$S$ finite. In turn, the scaling of $S$ in Fig. \ref{fig13}(b)
for increasing $h$ remains an open question.

\section{Conclusions and Speculations\label{sec:Conclusion}}

\begin{figure}
\centering{}\includegraphics[width=0.95\columnwidth]{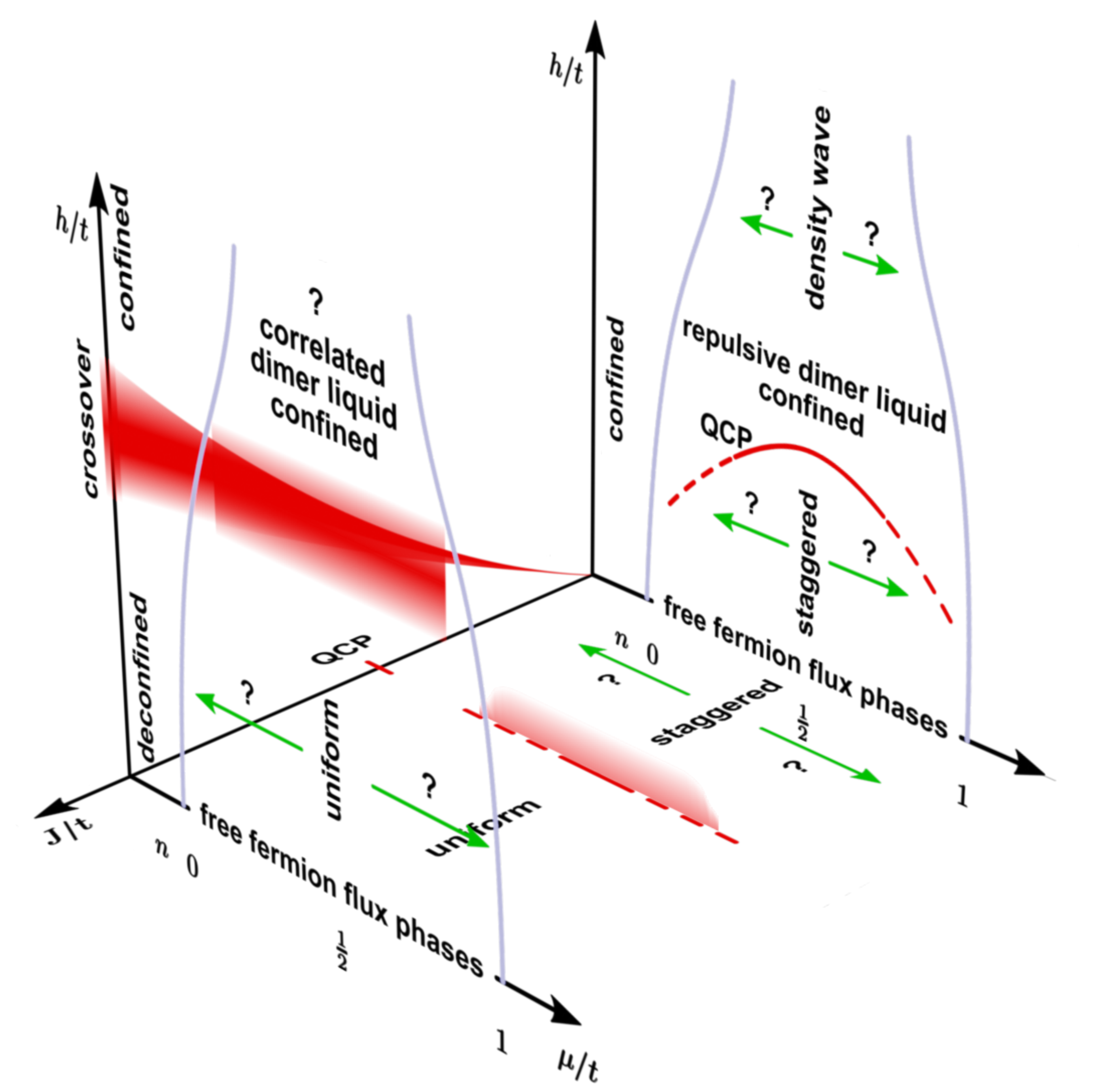}\caption{\label{fig:14}(Color online) Sketch of quantum phases of spinless
fermions in a $\mathbb{Z}_{2}$ gauge theory on the triangular ladder
resulting from this work. See Sec. \ref{sec:Conclusion} for details.}
\end{figure}

In conclusion, a study of the quantum phase diagram of spinless fermions
coupled to a constrained $\mathbb{Z}_{2}$ gauge theory on a triangular
ladder has been presented. Superficially the physics is similar to
that on other lattice structures, but the details are very different.
Simplifying the notation by $t_{i}\rightarrow t$, three dimensionless
parameters, filling ($\mu/t$), magnetic energy ($J/t$), and confinement
strength or electric coupling ($h/t$) control the overall behavior.
To summarize, a very rough and incomplete cartoon of this 3D space
is depicted in Fig. \ref{fig:14} for $J,h,\mu>0$, studied here.

For any finite $J/t$ and $h/t$, the system displays three phases
versus $\mu/t$, i.e., two pure gauge theories and one partially doped,
or filled regime. The latter is bounded by the curved lines in the
$\mu,h$-planes in Fig. \ref{fig:14}, symbolizing the band edges.
The even and odd pure gauge theories, left and right of these edges,
are strongly influenced by the triangular ladder structure and differ
from those on the square lattice. On the ladder and in the static
case at $h=0$, even and odd theories are unitarily equivalent and
for uniform electric coupling, $h>0$, confinement occurs by a crossover,
rather than by a QPT. Critical behavior can however be enforced using
an electric coupling confined to the rungs. The deconfinement-confinement
crossover is indicated by the red shaded wedge on the $J,h$-plane
in Fig. \ref{fig:14}. Obviously, as $J\rightarrow0$, any finite
$h$ implies immediate confinement. Topological order is not a meaningful
concept on the ladder because of the open boundary conditions transverse
to it, however, the ground states of the static pure gauge theories
still comprise a twofold degenerate quantum loop-gas.

For partial filling, two cases have been focused on, i.e., regions
of chemical potentials close to half filling and low fermion densities.
Looking at the former case in a $J,h$-plane in Fig. \ref{fig:14},
three phases could be identified. For vanishing $h$, the interplay
between the kinetic energy and the Peierls factor stabilizes flux
phases. At small $J/t$, close to the origin of the $J,h$-plane,
the latter is a staggered flux phase. At half filling, this is a band
insulator with spontaneously broken translational invariance. This
is different from the square lattice, where a $\pi$-flux Dirac semimetal
arises. While not investigated here, it is tempting to speculate that
the staggered flux phase might also be stable slightly off half filling
and for not too large, but finite $h/t$. Sufficiently far away from
that region, other flux phases may emerge. This is symbolized by the
question marks in Fig. \ref{fig:14}. Increasing $J$ at $h=0$ leads
to a first-order transition into a homogeneous flux phase. This is
indicated by the label QCP on the $J/t$ axis. Again, while not analyzed,
it seems plausible that this transition is not confined close to $\mu,h=0$
only. I.e., a 2D surface extends out of the $J,\mu$-plane within
the region symbolized by the dashed line and semi-transparent red
area, on which such transitions may occur. Question marks indicate
once more that the range of validity of this speculation is unclear. 

Finally, increasing $h/t$ enough, confinement of the fermions will
set in. How this occurs in detail is a matter of current debate. At
$n=1/2$ and $J=0$, the present study finds a single critical point
versus $h/t$, i.e., on the red line on $\mu,h$-plane in Fig. \ref{fig:14}.
This is consistent with $\mathbb{Z}_{2}$ gauge theories comprising
spinless, as well as spinful fermions on the square lattice. Fig.
\ref{fig:14} also displays some speculative extension of this QCP
into a line.

At very low density and strong confinement, i.e., $h\gg t,J$, the
model maps onto a dilute gas of nearest-neighbor fermionic dimers.
Here, it was demonstrated that these dimers feature kinetic energy
and interactions, all starting at second order in $t$ and $J$. The
interactions can be either repulsive or attractive, depending on the
relative magnitudes of $t$ and $J$. Due to this lack of a small
parameter, an analysis of the dimer gas remains an open question.
This situation is again different from the square lattice case, where
the attraction is of first order in $J$, allowing for simplifications
into a resonating dimer model. Nevertheless, at $J=0$, the confined
dimers are found to be purely repulsive. This suggests that at finite
doping density wave states can occur. Indeed, consistent with similar
findings on the square lattice, the present study has found incompressible
density waves for sufficiently large $h/t$ at half filling. This
refers to the third of the three phases, uncovered in this study at
$n=1/2$, and is indicated in the $\mu,h$-plane in Fig. \ref{fig:14}
at elevated $h/t$. Question marks label that the stability and commensuration
of such phases versus $\mu$ are open questions.

Finally, this study has provided global scans of the quantum phases
also at intermediate coupling. Yet, the details of the physics in
the region labeled 'correlated dimer liquid' at finite $J/t$, $h/t$,
and $\mu/t$ on the upper front plane in Fig. \ref{fig:14} are not
settled. If BEC, or BCS correlations, or phase separation can occur
on the triangular ladder, remains to be analyzed.

\begin{acknowledgments}
\emph{Acknowledgments}: Helpful communications with U. Borla, L. Janssen,
L. B. Jeevanesan, and S. Moroz are gratefully acknowledged. A critical
reading has been performed by A. Schwenke and E. Wagner. This work
has been supported in part by the DFG through project A02 of SFB 1143
(project-id 247310070). Kind hospitality of the PSM, Dresden, is acknowledged.
Initiation of this research was supported in part by the National
Science Foundation under Grant No. NSF PHY-1748958. MPS calculations
were performed using the TeNPy Library (version 0.8.4$-$0.9.0) \citep{Hauschild2018}.
COVID19 lockdowns are acknowledged, inducing single author's work.
\end{acknowledgments}

\vspace{15mm}

\appendix

\section{Mapping to pure spin model\label{app:A}}

In this section, the gauge theory with fermions on the triangular
ladder is mapped to a pure spin model which has only $4$ instead
of $8$ states per triangle. The new spin degrees of freedom are gauge-invariant
and the gauge constraint is satisfied by construction, i.e., the pure
spin model acts only in the physical subspace of zero gauge charge.
Variants of this approach have been described for 1D \citep{Cobanera2013,Radicevic2018,Borla2020,Borla2021a}
and 2D \citep{Borla2021} systems in the literature. The details are
specific to the particular lattice considered. Therefore, in the following,
this mapping is revisited for the triangular ladder.

\subsection{Gauge-invariant spin operators \label{subsec:A1}}

\begin{figure}
\centering{}\includegraphics[width=0.7\columnwidth]{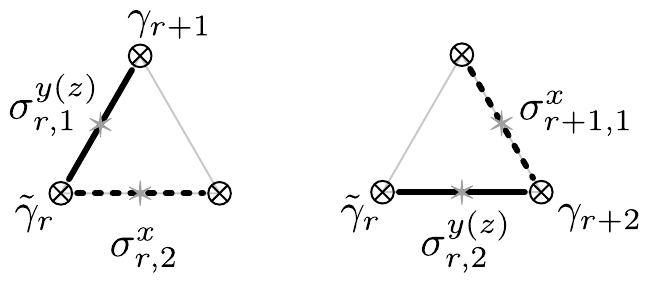}\caption{\label{fig14}Arrangement of 'dangling' $\sigma^{x}$ spin in Eq.
(\ref{eq:9}). }
\end{figure}

To begin, Majorana fermions $\gamma_{r}=c_{r}^{\dagger}+c_{r}^{\phantom{\dagger}}$
and $\tilde{\gamma}_{r}=i(c_{r}^{\dagger}-c_{r}^{\phantom{\dagger}})$
are introduced on the original fermion sites, with $\{\gamma_{r},\gamma_{r}\}=\{\tilde{\gamma}_{r},\tilde{\gamma}_{r}\}=2$,
$\{\gamma_{r},\tilde{\gamma}_{r}\}=0$, $\gamma_{r}^{2}=\tilde{\gamma}_{r}^{2}=1$,
and $\{\gamma_{r},\ptilde{\gamma}_{s}\}=0;\,\forall r\neq s$. Using
these, new spin operators $X,Y$, and $Z$ are defined on the sites
$(r,j)$ of the dual lattice by
\begin{align}
X_{r,j} & =\sigma_{r,j}^{x}\nonumber \\
Y(Z)_{r,1} & =-i\tilde{\gamma}_{r}\sigma_{r,1}^{y(z)}\gamma_{r+1}\sigma_{r,2}^{x}\label{eq:9}\\
Y(Z)_{r,2} & =-i\tilde{\gamma}_{r}\sigma_{r,2}^{y(z)}\gamma_{r+2}\sigma_{r+1,1}^{x}\,.\nonumber 
\end{align}
Using the transformation of $\sigma_{r}^{x,y,z}$ and $c_{r}^{(\dagger)}$
under the $\mathbb{Z}_{2}$ generator $G_{s}$, it is clear that $G_{s}(X,Y,Z)_{r,j}G_{s}=(X,Y,Z)_{r,j};\,\forall s$
, i.e., the new spins are indeed gauge-invariant. The 'dangling' $\sigma^{x}$
operator on $Y$ and $Z$, is peculiar to this mapping. In strictly
1D chain models \citep{Cobanera2013,Radicevic2018,Borla2020,Borla2021a}
it is absent. In 2D \citep{Borla2021} and for the present triangular
ladder it is required to obtain the proper spin algebra. Yet, this
latter requirement does not fix the placement of the dangling $\sigma_{r,j}^{x}$
uniquely and Eq. (\ref{eq:9}) is simply a convenient choice. The
arrangement is depicted in Fig. \ref{fig14}. Before using Eq. (\ref{eq:9})
in actual calculations, a detail is noted which may easily sink into
oblivion, namely, that the elements $\sigma^{x,y,z}$ of the original
Pauli algebra commute with all Majorana fermions by definition, however,
the new $Y$ and $Z$ certainly do \emph{not}.

To check the spin algebra, its on-site behavior is considered first.
Obviously, $X_{r,j}^{2}=Y_{r,j}^{2}=Z_{r,j}^{2}=1$. Moreover
\begin{align}
[X_{r,1},Y_{r,1}]= & -i\sigma_{r,1}^{x}\tilde{\gamma}_{r}\sigma_{r,1}^{y}\gamma_{r+1}\sigma_{r,2}^{x}\nonumber \\
 & +i\tilde{\gamma}_{r}\sigma_{r,1}^{y}\gamma_{r+1}\sigma_{r,2}^{x}\sigma_{r,1}^{x}\nonumber \\
= & -i\tilde{\gamma}_{r}[\sigma_{r,1}^{x},\sigma_{r,1}^{y}]\gamma_{r+1}\sigma_{r,2}^{x}\nonumber \\
= & \phantom{{-}}2 i Z_{r,1}\,,\label{eq:10}
\end{align}
and an identical relation for $[X_{r,2},Y_{r,2}]=2 i Z_{r,2}$, as well
as the cyclic equivalents $[Z_{r,j},X_{r,j}]=2 i Y_{r,j}$. Moreover,
\begin{align}
[Y_{r,1},Z_{r,1}]= & -\tilde{\gamma}_{r}\sigma_{r,1}^{y}\gamma_{r+1}\sigma_{r,2}^{x}\tilde{\gamma}_{r}\sigma_{r,1}^{z}\gamma_{r+1}\sigma_{r,2}^{x}\nonumber \\
 & +\tilde{\gamma}_{r}\sigma_{r,1}^{z}\gamma_{r+1}\sigma_{r,2}^{x}\tilde{\gamma}_{r}\sigma_{r,1}^{y}\gamma_{r+1}\sigma_{r,2}^{x}\nonumber \\
= & +\sigma_{r,1}^{y}\sigma_{r,2}^{x}\sigma_{r,1}^{z}\sigma_{r,2}^{x}-\sigma_{r,1}^{z}\sigma_{r,2}^{x}\sigma_{r,1}^{y}\sigma_{r,2}^{x}\nonumber \\
= & \phantom{{+\}}}[\sigma_{r,1}^{y},\sigma_{r,1}^{z}]=2 i X_{r,1}\,,\label{eq:11}
\end{align}
and identically, $[Y_{r,2},Z_{r,2}]=2 i X_{r,2}$.

Second, off-site commutation relations between the new spins on dual
sites $(r,j)$ and $(s,m)$ are considered, corresponding to two nearest-neighbor
links which share just one Majorana fermion. Two cases arise. Either
all original spins reside on different dual sites, or two spins are
from identical links. An example for the former is
\begin{align}
[Y_{r,1},Y_{r-2,2}]= & -\tilde{\gamma}_{r}\sigma_{r,1}^{y}\gamma_{r+1}\sigma_{r,2}^{x}\tilde{\gamma}_{r-2}\sigma_{r-2,2}^{y}\gamma_{r}\sigma_{r-1,1}^{x}\nonumber \\
 & +\tilde{\gamma}_{r-2}\sigma_{r-2,2}^{y}\gamma_{r}\sigma_{r-1,1}^{x}\tilde{\gamma}_{r}\sigma_{r,1}^{y}\gamma_{r+1}\sigma_{r,2}^{x}\nonumber \\
= & \phantom{+}\tilde{\gamma}_{r-2}\gamma_{r+1}\{\gamma_{r},\tilde{\gamma}_{r}\}\sigma_{r-2,2}^{y}\sigma_{r-1,1}^{x}\sigma_{r,1}^{y}\sigma_{r,2}^{x}\nonumber \\
= & \phantom{+}0\,,\label{eq:12}
\end{align}
i.e., the Majorana algebra renders the commutator proper. To appreciate
the action of the dangling $\sigma^{x}$ operators, nearest neighbor
commutators for the second case are now evaluated 
\begin{align}
[Y_{r,1},Y_{r,2}]= & -\tilde{\gamma}_{r}\sigma_{r,1}^{y}\gamma_{r+1}\sigma_{r,2}^{x}\tilde{\gamma}_{r}\sigma_{r,2}^{y}\gamma_{r+2}\sigma_{r+1,1}^{x}\nonumber \\
 & +\tilde{\gamma}_{r}\sigma_{r,2}^{y}\gamma_{r+2}\sigma_{r+1,1}^{x}\tilde{\gamma}_{r}\sigma_{r,1}^{y}\gamma_{r+1}\sigma_{r,2}^{x}\nonumber \\
= & +\gamma_{r+1}\gamma_{r+2}\sigma_{r,1}^{y}\sigma_{r,2}^{x}\sigma_{r,2}^{y}\sigma_{r+1,1}^{x}\nonumber \\
 & -\gamma_{r+2}\gamma_{r+1}\sigma_{r,2}^{y}\sigma_{r+1,1}^{x}\sigma_{r,1}^{y}\sigma_{r,2}^{x}\nonumber \\
= & \phantom{{+}}\gamma_{r+1}\gamma_{r+2}\sigma_{r,1}^{y}\sigma_{r+1,1}^{x}\{\sigma_{r,2}^{x},\sigma_{r,2}^{y}\}=0\,.\label{eq:13}
\end{align}
This shows that the dangling $\sigma^{x}$ operators are necessary
to fix the commutator for those cases where the Majorana fermions
which are shared by both new spin operators are of the type $\gamma_{r}\gamma_{r}$
or $\tilde{\gamma}_{r}\tilde{\gamma}_{r}$, instead of $\gamma_{r}\tilde{\gamma}_{r}$.
This also clarifies why dangling $\sigma^{x}$ operators only have
to be introduced on lattice graphs which are not of strict chain-type.

Similar to Eqs. (\ref{eq:12}) and (\ref{eq:13}), it is simple to
show that all commutators of $X,Y$, and $Z$ operators on nearest-neighbor
links commute. On dual sites which are farther apart, the new spins
commute trivially, because all operators from the right-hand side
of Eq. (\ref{eq:9}) are different and the number of Majoranas to
commute is even.

In conclusion, Eq. (\ref{eq:9}) does indeed represent a gauge-invariant
spin algebra.

\subsection{Pure spin-model \label{subsec:A2}}

To begin, the kinetic energy of the fermions from Eq. (\ref{eq:2})
is transformed. This is done in several steps. First, in terms of
the Majorana fermions
\begin{align}
 & -\sum_{r,j=1,2}t_{j}(c_{r+j}^{\dagger}\sigma_{r,j}^{z}c_{r}^{\phantom{\dagger}}+c_{r}^{\dagger}\sigma_{r,j}^{z}c_{r+j}^{\phantom{\dagger}})=\nonumber \\
 & \phantom{{-}}\frac{1}{2}\sum_{r,j=1,2}t_{j}(i\tilde{\gamma}_{r}\sigma_{r,j}^{z}\gamma_{r+j}-i\gamma_{r}\sigma_{r,j}^{z}\tilde{\gamma}_{r+j})=\nonumber \\
 & \phantom{{-}}\frac{-1}{2}\sum_{r,j=1,2}t_{j}(\delta_{j,1}Z_{r,1}X_{r,2}+\delta_{j,2}Z_{r,2}X_{r+1,1}\nonumber \\
 & \phantom{{aaaaaaaaaaaaa}}+i\gamma_{r}\sigma_{r,j}^{z}\tilde{\gamma}_{r+j})=(\star)\,,\label{eq:14}
\end{align}
Where on the third line, Eq. (\ref{eq:9}) has been inserted. On the
last line, the $\sim$ labeling of the Majorana fermions is unfavorable
for direct insertion of the new spin operators. However, the gauge
constraint can be invoked to cure this. Namely, with $i\tilde{\gamma}_{r}\gamma_{r}(-)^{n_{r}}=(1-2n_{r})(-)^{n_{r}}=1$,
Eq. (\ref{eq:5}) with $G_{r}=1$, can be rewritten as

\begin{equation}
1=i\tilde{\gamma}_{r}\gamma_{r}\prod_{b\in S_{r}}\sigma_{b}^{x}=i\tilde{\gamma}_{r}\gamma_{r}\prod_{b\in S_{r}}X_{b}\equiv i\tilde{\gamma}_{r}\gamma_{r}\mathbb{A}_{r}\,.\label{eq:15}
\end{equation}
To ease the notation and because of the first line of Eq. (\ref{eq:9}),
as well as because of the definition of $A_{r}$ from Eq. (\ref{eq:5}),
the symbol $\mathbb{A}_{r}$ is introduced, which is mathematically
identical to $A_{r}$, and meant only to denote the relabeling $\sigma_{r,j}^{x}\rightarrow X_{r,j}$.
The unity (\ref{eq:15}) can be inserted as follows
\begin{align}
 & i\gamma_{r}\sigma_{r,j}^{z}\tilde{\gamma}_{r+j}=i\gamma_{r}\sigma_{r,j}^{z}\tilde{\gamma}_{r+j}G_{r}G_{r+j}=\nonumber \\
 & -i\gamma_{r}\sigma_{r,j}^{z}\tilde{\gamma}_{r+j}\tilde{\gamma}_{r}\gamma_{r}\tilde{\gamma}_{r+j}\gamma_{r+j}\mathbb{A}_{r}\mathbb{A}_{r+j}=\nonumber \\
 & i\sigma_{r,j}^{z}\tilde{\gamma}_{r}\gamma_{r+j}\mathbb{A}_{r}\mathbb{A}_{r+j}=\nonumber \\
 & i\tilde{\gamma}_{r}\sigma_{r,j}^{z}\gamma_{r+j}\mathbb{A}_{r}\mathbb{A}_{r+j}\label{eq:16}
\end{align}
where on the 2nd line the Majoranas from the gauge constraint are
labeled such as to compensate the improperly labeled ones from the
hopping. This trick can be applied to arbitrary Majorana products
in order to relabel the $\sim$ accents at the expense of introducing
additional star operators $\mathbb{A}_{r}$. With Eq. (\ref{eq:16})
\begin{align}
(\star)=-\frac{1}{2}\sum_{r} & \left[t_{1}Z_{r,1}X_{r,2}(1-\mathbb{A}_{r}\mathbb{A}_{r+1})\right.\nonumber \\
 & \left.+t_{2}Z_{r,2}X_{r+1,1}(1-\mathbb{A}_{r}\mathbb{A}_{r+2})\right]\,.\label{eq:17}
\end{align}
Because of the gauge constraint $(-)^{n_{r}}A_{r}=1$, the terms $(1-\mathbb{A}_{r}\mathbb{A}_{u})/2\equiv P_{ru}$
serve as \emph{projectors} \citep{Borla2021}, which guarantee that
the hopping process, encoded in the preceding transformed expression,
can only occur between sites $r,u$ of different fermion parity, i.e.,
such that no double occupancy is generated.

The transformation of the density for the chemical potential term
can be adopted directly from ref. \citep{Borla2021}, using that because
of the $\mathbb{Z}_{2}$ Gau{\ss} law
\begin{equation}
2n_{r}=1-\mathbb{A}_{r}\,.\label{eq:18}
\end{equation}
Finally, the transformation of the magnetic field energy needs to
be considered. From Eqs. (\ref{eq:3},\ref{eq:4})
\begin{equation}
B_{r}=\sigma_{r,1}^{z}\sigma_{r,2}^{z}\sigma_{r+1,1}^{z}\,,\label{eq:19}
\end{equation}
where $r$ refers to the lower(upper) left corner of the plaquette
for up(down)ward pointing triangles. With Eq. (\ref{eq:9}) this reads
\begin{align}
B_{r}= & i\gamma_{r+1}\tilde{\gamma}_{r+1}Z_{r,1}X_{r,2}Z_{r,2}X_{r+1,1}Z_{r+1,1}X_{r+1,2}\nonumber \\
= & -\mathbb{A}_{r+1}Z_{r,1}X_{r,2}Z_{r,2}X_{r+1,1}Z_{r+1,1}X_{r+1,2}\,,\label{eq:20}
\end{align}
where, again, the unity (\ref{eq:15}) has been used to eliminate
the remaining Majorana fermions.

Because of the spin algebra, expressions like (\ref{eq:20}), or those
in (\ref{eq:17}), comprising stars, may allow for additional reduction.
E.g., $B_{r}$ simplifies to
\begin{align}
B_{r}= & -X_{r-1,2}X_{r,1}X_{r+1,1}X_{r+1,2}\nonumber \\
 & \phantom{aaa}Z_{r,1}X_{r,2}\,\,Z_{r,2}X_{r+1,1}\,\,Z_{r+1,1}X_{r+1,2}\nonumber \\
= & X_{r-1,2}Y_{r,1}Y_{r,2}Z_{r+1,1}\,.\label{eq:21}
\end{align}
While this section shows that the general principles of the mapping
for the triangular ladder are identical to those for the square lattice
\citep{Borla2021}, the preceding equation also highlights that the
details are different. I.e., while on the square lattice the magnetic
field energy turns into products of plaquettes and stars, for the
triangular ladder this is not so.

From Eqs. (\ref{eq:9})-(\ref{eq:21}), the chain model of Eqs. (\ref{eq:6a},\ref{eq:7a})
can be read off after some simple re-indexing.

\end{document}